\documentclass[conference]{IEEEtran}
\usepackage{amsmath,amssymb,amsfonts}
\usepackage{amsthm}
\usepackage{wasysym}
\usepackage{enumitem}

\usepackage{todonotes}
\usepackage{multicol}
\usepackage{wrapfig}

\usepackage{marginnote}
\usepackage{framed}
\newtheorem{example}{Example}
\newtheorem{definition}{Definition}

\usepackage{amsmath,amssymb,amsfonts}
\usepackage{amsthm}
\usepackage{wasysym}
\usepackage{varwidth}
\usepackage{algorithm}
\usepackage{url}
\usepackage[noend]{algpseudocode}
\usepackage{array}

\usepackage[clock]{ifsym}

\usepackage{fontawesome}

\usepackage{multirow}

\usepackage{pifont}

\usepackage{caption}
\usepackage{subcaption}

\usepackage{graphicx}
\usepackage[export]{adjustbox}

\usepackage{balance}

\usepackage{xcolor}

\usepackage{enumitem}

\newcommand{\N}{\mathbb{N}}

\newcommand{\incell}[2]{\begin{tabular}{@{}#1@{}}#2\end{tabular}}

\newcommand{\npc}{\text{NP-complete}}

\newcommand{\nphardness}{\text{NP-hardness}}

\newcommand{\ok}{{\sf ok}}

\newcommand{\Key}{{\sf K}}
\newcommand{\Val}{{\sf V}}

\newcommand{\h}{\mathcal{H}}
\newcommand{\opset}{\mathit{O}}

\newcommand{\opvar}{\mathit{op}}
\newcommand{\readevent}{{\sf R}}

\newcommand{\writeevent}{{\sf W}}

\renewcommand{\ae}{\mathcal{A}}
\newcommand{\rel}[1]{\xrightarrow{#1}}
\newcommand{\comp}{\;;\;}

\newcommand{\po}{{\sf po}}

\newcommand{\so}{\textup{\textsc{SO}}}
\newcommand{\vis}{\textup{\textsc{VIS}}}
\newcommand{\ar}{\textup{\textsc{AR}}}

\newcommand{\intaxiom}{\textup{\textsc{Int}}}
\newcommand{\extaxiom}{\textup{\textsc{Ext}}}
\newcommand{\sessionaxiom}{\textup{\textsc{Session}}}
\newcommand{\prefixaxiom}{\textup{\textsc{Prefix}}}
\newcommand{\noconflictaxiom}{\textup{\textsc{NoConflict}}}
\newcommand{\si}{\textsc{SI}}

\newcommand{\T}{\mathcal{T}}

\renewcommand{\H}{\mathcal{H}}

\newcommand{\SO}{\textsf{SO}}

\newcommand{\blue}[1]{\textcolor{blue}{#1}}

\newcommand{\brown}[1]{\textcolor{brown}{#1}}

\newcommand{\violet}[1]{\textcolor{violet}{#1}}
\colorlet{lightteal}{teal!60}
\newcommand{\lightteal}[1]{\textcolor{lightteal}{#1}}

\newcommand{\set}[1]{\{#1\}}
\newcommand{\tuple}[1]{\langle#1\rangle} % or using (#1)?
\newcommand{\domain}{\textsc{Dom}}

\newcommand{\code}[2]{line~#1:#2}
\newcommand{\codes}[3]{lines~#1:#2 -- #1:#3}
\newcommand{\hStatex}[0]{\vspace{4pt}}

\newcommand{\keysvar}{\mathit{keys}}

\newcommand{\TS}{{\sf TS}}
\newcommand{\ts}{{\sf ts}}
\newcommand{\tsvar}{\mathit{ts}}
\renewcommand{\ae}{\mathcal{A}}

\newcommand{\tool}{\textup{\textsc{Chronos}}}
\newcommand{\onlinetool}{\textup{\textsc{Aion}}}

\newcommand{\starttxn}{\textsc{Start}}
\newcommand{\readtxn}{\textsc{Read}}
\newcommand{\writetxn}{\textsc{Write}}
\newcommand{\committxn}{\textsc{Commit}}
\newcommand{\timestamprequest}{\textsc{Request}}

\newcommand{\buffer}{\mathit{buffer}}
\newcommand{\timeoracle}{\mathcal{O}}

\renewcommand{\log}{\textsf{log}}
\newcommand{\committed}{\textsf{committed}}
\newcommand{\aborted}{\textsf{aborted}}

\newcommand{\starttime}{\mathit{start\_ts}}
\newcommand{\committime}{\mathit{commit\_ts}}

\newcommand{\cts}{\mathit{cts}}
\newcommand{\beforets}{\widehat{\tsvar}}

\newcommand{\beforects}{\widehat{\cts}}

\newcommand{\tid}{\mathit{tid}}
\newcommand{\sid}{\mathit{sid}}
\newcommand{\sno}{\mathit{sno}}
\newcommand{\ops}{\mathit{ops}}

\newcommand{\intval}{\textsf{int\_val}}
\newcommand{\extval}{\textsf{ext\_val}}

\newcommand{\lastsnoinsession}{\textsf{last\_sno}}
\newcommand{\lastctsinsession}{\textsf{last\_cts}}
\newcommand{\frontier}{\textsf{frontier}}
\newcommand{\frontierts}{\textsf{frontier\_{ts}}}
\newcommand{\ongoing}{\textsf{ongoing}}
\newcommand{\ongoingts}{\textsf{ongoing\_{ts}}}

\newcommand{\gc}{\textsc{GarbageCollect}}

\newcommand{\htread}{\mathit{ht\_read}}

\newcommand{\TID}{\textsf{TID}}
\newcommand{\SID}{\textsf{SID}}
\newcommand{\nulltxn}{\bot_{T}}
\newcommand{\nullval}{\bot_{v}}
\newcommand{\nullts}{\bot_{\mathit{ts}}}

\newcommand{\checksi}{\textsc{CheckSI}}
\newcommand{\onlinechecksi}{\textsc{OnlineCheckSI}}

\newcommand{\wkey}{\mathit{wkey}}

\newcommand{\frontiertsstyle}[1]{\colorbox{blue!30}{$#1$}}
\newcommand{\ongoingtsstyle}[1]{\colorbox{green!30}{$#1$}}
\newcommand{\diskstyle}[1]{\boxed{\colorbox{teal!30}{$#1$}}}
\newcommand{\recheckextoptstyle}[1]{\underline{#1}}
\newcommand{\gcstyle}[1]{\boxed{#1}}

\newcommand{\startstyle}[1]{\blue{#1}}
\newcommand{\commitstyle}[1]{\brown{#1}}
\newcommand{\clockstyle}{\violet{\scriptsize \StopWatchStart}{}}
\newcommand{\loadstyle}{\lightteal{\faUpload}} % \leftturn
\newcommand{\savestyle}{\lightteal{\faDownload}}

\newcommand{\circleone}{\ding{182}}
\newcommand{\circletwo}{\ding{183}}

\newcommand{\circlefour}{\ding{185}}

\newcommand{\circlesix}{\ding{187}}
\newcommand{\circleseven}{\ding{188}}
\newcommand{\circleeight}{\ding{189}}
\newcommand{\circlenine}{\ding{190}}
\newcommand{\circleten}{\ding{191}}

\AtBeginDocument{%
  \providecommand\BibTeX{{%
    Bib\TeX}}}

\def\BibTeX{{\rm B\kern-.05em{\sc i\kern-.025em b}\kern-.08em
    T\kern-.1667em\lower.7ex\hbox{E}\kern-.125emX}}
\begin{document}

% \sloppypar
% \input{sections/cover_letter}

% \title{\tool{} and \onlinetool: Offline and Online Timestamp-based Checking of Snapshot Isolation in Database Systems
% }
\title{Online Timestamp-based Transactional Isolation Checking of Database Systems (Extended Version)}

\author{
\IEEEauthorblockN{
$\text{Hexu Li}^{1}$,
$\text{Hengfeng Wei}^{1*}$,
$\text{Hongrong Ouyang}^{2}$, 
$\text{Yuxing Chen}^{3}$,
$\text{Na Yang}^{1}$,
$\text{Ruohao Zhang}^{4}$,
$\text{Anqun Pan}^{3}$
}
\IEEEauthorblockN{
$\textit{State Key Laboratory for Novel Software Technology, Nanjing University}^1$,\\
$\textit{OceanBase, Ant Group}^2$,
$\textit{Tencent Inc.}^3$,
$\textit{The Chinese University of Hong Kong}^4$
}
\IEEEauthorblockN{
\{hexuli, nayang\}@smail.nju.edu.cn,
hfwei@nju.edu.cn,
ouyanghongrong.oyh@oceanbase.com,\\
\{axingguchen, aaronpan\}@tencent.com,
ruohaozhang@link.cuhk.edu.hk
}
}

\maketitle

% \begin{abstract}
% This document is a model and instructions for \LaTeX.
% This and the IEEEtran.cls file define the components of your paper [title, text, heads, etc.]. *CRITICAL: Do Not Use Symbols, Special Characters, Footnotes, 
% or Math in Paper Title or Abstract.
% \end{abstract}

% abstract.tex

\begin{abstract}
    Serializability (SER) and snapshot isolation (SI) are 
    widely used transactional isolation levels in database systems.
    The isolation checking problem asks
    whether a given execution history of a database system
    satisfies a specified isolation level.
    However, existing SER and SI checkers,
    whether traditional black-box checkers 
    or recent timestamp-based white-box ones,
    operate offline and require the entire history to be available to construct a dependency graph,
    making them unsuitable for continuous and ever-growing histories.

    This paper addresses online isolation checking
    by extending the timestamp-based isolation checking approach
    to online settings.
    Specifically, we design \tool{}, an efficient timestamp-based
    offline SI checker.
    \tool{} is incremental and avoids constructing
    a start-ordered serialization graph for the entire history,
    making it well-suited for online scenarios.
    We further extend \tool{} into an online SI checker, \onlinetool,
    addressing several key challenges unique to online settings.
    Additionally, we develop \onlinetool-SER for online SER checking.
    Experiments highlight that \tool{} processes offline histories
    with up to one million transactions in seconds,
    greatly outperforming existing SI checkers.
    Furthermore, \onlinetool{} and \onlinetool-SER
    sustain a throughput of approximately 12K transactions per second,
    demonstrating their practicality for online isolation checking.
\end{abstract}


% \begin{abstract}
	% The snapshot isolation (SI) checking problem,
	% which determines whether a given database execution satisfies SI,
	% is fundamental to database testing.
	% Most existing SI checkers treat the database system as a black-box,
	% lacking insight into the underlying SI implementation
	% and particularly the actual execution order of transactions.
	% Consequently, these tools often face computational complexity issues
	% and may be restricted to specific data types in database executions.
	% Moreover, they are offline checkers with limited scalability.

	% To overcome these drawbacks,
	% we propose a novel \emph{timestamp-based} approach to SI checking.
	% This approach requires database executions to include
	% the start and commit timestamps of each transaction,
	% making it applicable regardless of the data types used.
	% By leveraging these timestamps,
	% we can infer the actual execution order of transactions
	% and determine the snapshot of each transaction.
	% We have developed both offline and online
	% highly efficient SI checking algorithms,
	% named \tool{} and \onlinetool, respectively.
	% Experimental results demonstrate that
	% \tool{} can handle offline histories
	% containing up to a million of transactions within tens of seconds,
	% while existing black-box SI checkers take several hours
	% or cannot handle such large histories at all.
	% Moreover, \onlinetool{} can handle online transactional workloads
	% with a {sustained} throughput of approximately
	% 12K transactions per second.
% \end{abstract}

% \begin{IEEEkeywords}
% component, formatting, style, styling, insert.
% \end{IEEEkeywords}

\pagenumbering{arabic}
\pagestyle{plain}
\setcounter{section}{0}
% intro.tex

%%%%%%%%%%%%%%%%%%%%%%%%%%%%%%
\section{Introduction}
\label{section:intro}

% \bfit{Transactions and Isolation Levels.}
Database transactions provide an ``all-or-none'' abstraction
and isolate concurrent computations on shared data,
greatly simplifying client programming.
Serializability (SER)~\cite{SER:JACM1979,Bernstein:Book1986},
the gold standard for transactional isolation,
ensures that all transactions appear to execute sequentially.
Prominent database systems offering SER include
PostgreSQL~\cite{PostgreSQL}, CockroachDB~\cite{CockroachDB}, and YugabyteDB~\cite{YugabyteDB}.
For improved performance~\cite{HAT:VLDB2013},
% Ensuring SER, particularly in a distributed setting,
% is inevitably expensive.
many databases implement snapshot isolation (SI)~\cite{CritiqueANSI:SIGMOD1995},
a widely adopted weaker isolation level.
Examples include YugabyteDB,
and Dgraph~\cite{dgraph}, WiredTiger~\cite{WiredTiger-Transaction},
Google's Percolator~\cite{Percolator:OSDI2010},
MongoDB~\cite{MongoDB}, and TiDB~\cite{TiDB}.
% including both classic centralized databases
% (e.g., PostgreSQL and WiredTiger~\cite{WiredTiger-Transaction})
% and emerging distributed databases
% (e.g., Google's Percolator~\cite{Percolator:OSDI2010},
% MongoDB~\cite{MongoDB}, TiDB~\cite{TiDB}, YugabyteDB,
% and Dgraph~\cite{dgraph}).
% and Apache Omid~\cite{Omid:FAST2017}.
% SI prevents clients from various undesired data anomalies including fractured reads, causality violations,
% lost update, and long fork~\cite{AnalysingSI:JACM2018}.
% , but it allows write skew anomalies.

% \paragraph*{Black-box SI Checking}
Implementing databases correctly, however, is notoriously challenging,
and many fail to deliver SER or SI as claimed~\cite{PostgreSQL-Jepsen, MongoDB-Jepsen, TiDB-Jepsen, Cobra:OSDI2020, Complexity:OOPSLA2019, CockroachDB-Jepsen}.
The isolation checking problem of determining whether a given execution history satisfies a specified isolation level
is fundamental to database testing.
\emph{Black-box} checkers for SER and SI,
such as dbcop~\cite{Complexity:OOPSLA2019},
Cobra~\cite{Cobra:OSDI2020}, Elle~\cite{Elle:VLDB2020},
PolySI~\cite{PolySI:VLDB2023}, and Viper~\cite{Viper:EuroSys2023},
have been developed.
% In {black-box} settings, a history consists of a set of transactions
% issued by clients to database,
% and typically records the transactions' identifiers,
% the keys they access, and the corresponding values.
% In particular, it lacks the information of the actual execution order
% of these transactions.
However, due to the \nphardness{} of SER and SI checking in black-box settings~\cite{SER:JACM1979, Complexity:OOPSLA2019, DBLP:journals/pvldb/GuLXWCB24},
these checkers often exhibit high computational complexity
and struggle to scale to large histories
containing tens of thousands of transactions.
Recent advances propose timestamp-based \emph{white-box} checkers~\cite{VerifyingMongoDB:arXiv2022, SER-VO:ETH2021, Clark:EuroSys2024},
which leverage additional transaction start and commit timestamps
to infer execution order and read views. 
Particularly, Emme-SER and Emme-SI~\cite{Clark:EuroSys2024}
enable polynomial-time verification for SER and SI, respectively, 
alleviating scalability challenges in large histories.

\paragraph*{Problem}
To our knowledge, existing approaches are all \emph{offline} checkers,
requiring the entire history to be available before checking.
However, in practice, it is often desirable to perform isolation checking online,
i.e., on continuous and ever-growing history.
While Cobra~\cite{Cobra:OSDI2020} supports online SER checking,
it does not support online SI checking
and requires injecting ``fence transactions'' into client workloads
-- an approach often unacceptable in production.

\paragraph*{Our Work}
In this study, we tackle the problem of \emph{online} transactional isolation
checking for database systems
by extending the timestamp-based isolation checking approach to online settings.
On the one hand, online checkers must be highly efficient
to keep pace with the database throughput,
which generates a continuous and ever-growing history.
However, Emme-SI performs expensive graph construction and cycle detection
on the start-ordered serialization graph of the entire history~\cite{Adya:PhDThesis1999},
rendering it unsuitable for online use.
On the other hand, online checkers \emph{cannot} assume that
transactions are collected in ascending order
of their start or commit timestamps,
as asynchrony in database systems makes this infeasible.
This introduces three key challenges unique to online checkers,
which we address in our work.

\begin{itemize}
  \item First, the satisfaction or violation of isolation level rules
    for a transaction becomes \emph{unstable} due to asynchrony,
    making it impossible to assert correctness
    immediately upon transaction collection.
  \item Second, an incoming transaction $T$ with a smaller start timestamp
		may require \emph{re-checking} transactions 
		that commit after $T$ started.
        Efficient data structures are necessary to facilitate this re-checking process.
  \item Third, to ensure long-term feasibility and prevent unbounded memory usage,
      online checkers must \emph{recycle} data structures and transactions no longer needed.
      However, in the worst-case scenario,
      asynchrony may prevent the safe recycling of any data structures or transactions.
\end{itemize}

\paragraph*{Contributions}
Our approach supports both online SER and SI checking
and is general and adaptable to various data types used in generating histories.
In this work, we focus on online SI checking,
particularly for key-value histories and list histories.\footnote{
  While theoretically similar to online SER checking,
  online SI checking poses greater engineering challenges.}
Specifically, our contributions are as follows:

\begin{itemize}
	\item (Section~\ref{ss:chronos})
	  We design \tool, a highly efficient timestamp-based offline SI checker
      with a time complexity of $O(N \log N + M)$,
		where $N$ and $M$ represent the number of transactions and
		operations in the history, respectively.
        Unlike Emme-SI, \tool{} is incremental and
        avoids constructing the start-ordered serialization graph 
        for the entire history, making it well-suited for online checking.
	\item (Section~\ref{ss:aion})
	  We identify the challenges of online SI checking
        caused by inconsistencies between the execution
		and checking orders of transactions.
		To address these challenges,
		we extend \tool{} to support online checking,
        introducing a new checker named \onlinetool.
	\item (Sections~\ref{section:offline-experiments}
	  and \ref{section:online-experiments})
	  We evaluate \tool{} and \onlinetool{}
	  under a diverse range of workloads.
		% on database systems of different types.
		% including TiDB, MongoDB, YugabyteDB, and Dgraph.
		Results show that \tool{} can process offline histories
		with up to one million transactions in tens of seconds,
        significantly outperforming existing SI checkers,
        including Emme-SI,
        which either take hours or fail to handle such large histories. 
		Moreover, \onlinetool{} supports online transactional workloads 
		with a sustained throughput of approximately 12K transactions per second (TPS),
		with only a minor (approximately 5\%) impact
		on database throughput, primarily due to history collection.
    \item (Section~\ref{section:online-experiments})
      We also develop \onlinetool-SER, an online SER checker.
      Experimental results demonstrate that \onlinetool-SER supports
      online transactional workloads with a sustained throughput of approximately 12K TPS, greatly outperforming Cobra.
\end{itemize}
\section{Semantics of Snapshot Isolation}
\label{section:si-semantics}

In this section, we review both the operational and axiomatic semantics of SI.
In Section~\ref{section:ts-checker},
we design our timestamp-based SI checking algorithms
by relating these two semantics.

We consider a key-value store managing a set $\Key$ of keys
associated with values from $\Val$.\footnote{
	We assume an artificial value $\nullval \notin \Val$.}
We use $\readevent(k, v)$ to denote a read operation
that reads $v \in \Val$ from $k \in \Key$
and $\writeevent(k, v)$ to denote a write operation
that writes $v \in \Val$ to $k \in \Key$.

% si-op.tex

%%%%%%%%%%%%%%%%%%%%%%%%%%%%%%
\subsection{SI: Operational Semantics}
\label{ss:si-operational}

% si.tex

\small
\begin{algorithm}[t]
  \caption{Operational Semantics of Snapshot Isolation}
  \label{alg:si}
  % \begin{varwidth}[t]{0.46\textwidth}
  \begin{algorithmic}[1]
		\Statex $\log$: the log of committed transactions

		\hStatex
    \Procedure{\starttxn}{$T$}
      \label{line:proc-starttxn}
			\State $T.\starttime \gets \timestamprequest(\timeoracle)$
			  \label{line:starttxn-starttime}
			\State \Return \ok
    \EndProcedure

		\hStatex
    \Procedure{\writetxn}{$T, k, v$}
      \label{line:proc-writetxn}
			\State $T.\buffer \gets T.\buffer \circ \tuple{k, v}$
			  % \Comment{append $\tuple{k, v}$ to $T.\buffer$}
			  \label{line:writetxn-buffer}
			\State \Return \ok
    \EndProcedure

		\hStatex
    \Procedure{\readtxn}{$T, k$}
      \label{line:proc-readtxn}
			\State \Return value of $k$ from $T.\buffer$
			  and $\log$ as of $T.\starttime$
				\label{line:readtxn-value}
    \EndProcedure

		\hStatex
    \Procedure{\committxn}{$T$}
      \label{line:proc-committxn}
			\State $T.\committime \gets \timestamprequest(\timeoracle)$
			  \label{line:committxn-committime}
			\If{$T$ \emph{conflicts} with some concurrent transaction}
			  \label{line:committxn-conflict-checking}
			  \State \Return $\aborted$
					\label{line:aborttxn-status}
			\EndIf
			\State $\log \gets \log \circ \tuple{T.\buffer, T.\committime}$
				% \Comment{append $T$ to $\log$}
				\label{line:committxn-log}
			\State \Return $\committed$
			  \label{line:committxn-return-status}
    \EndProcedure
  \end{algorithmic}
  % \end{varwidth}
\end{algorithm}
\normalsize

SI was first defined in \cite{CritiqueANSI:SIGMOD1995}.
The original definition is operational:
it is describing how an implementation would work.
We specify SI by showing a high-level implementation of it
in Algorithm~\ref{alg:si}~\cite{Adya:PhDThesis1999, PSI:SOSP2011}.
Each procedure in Algorithm~\ref{alg:si} is executed atomically.
We assume a time oracle $\timeoracle$ which returns a unique timestamp upon request
and that the timestamps are totally ordered.
In the paper, we reference pseudocode lines using the format ``algorithm\#:line\#''.

When a transaction $T$ starts,
it is assigned a start timestamp $T.\starttime$
(\code{\ref{alg:si}}{\ref{line:starttxn-starttime}}).
The writes of $T$ are buffered in $T.\buffer$
(\code{\ref{alg:si}}{\ref{line:writetxn-buffer}}).
Under SI, the transaction $T$ reads data from
its own write buffer and the snapshot (stored in $\log$)
of \emph{committed} transactions
valid as of its start time $T.\starttime$
(\code{\ref{alg:si}}{\ref{line:readtxn-value}}).
When $T$ is ready to commit,
it is assigned a commit timestamp $T.\committime$
(\code{\ref{alg:si}}{\ref{line:committxn-committime}})
and allowed to commit if no other \emph{concurrent} transaction $T'$
(i.e., $[T'.\starttime, T'.\committime]$
and $[T.\starttime, T.\committime]$ overlap)
has already written data that $T$ intends to write
(\code{\ref{alg:si}}{\ref{line:committxn-conflict-checking}}).
This rule prevents the ``lost updates'' anomaly.
If $T$ commits, it appends its buffered writes,
along with its commit timestamp, to $\log$
(\code{\ref{alg:si}}{\ref{line:committxn-log}}).

% The start and commit timestamps of a transaction $T$
% can be wall time or logical time.
We require that for each transaction $T$, 
$T.\starttime \le T.\committime$.
\begin{align}
	T.\starttime \le T.\committime. 
	\label{eq:si-startts-committs}
\end{align}
Note that we allow $T.\committime = T.\starttime$,
which is possible for read-only transactions.

% si-op-axiomatic.tex

\begin{figure}[t]
	\centering
	\includegraphics[width = 0.35\textwidth]{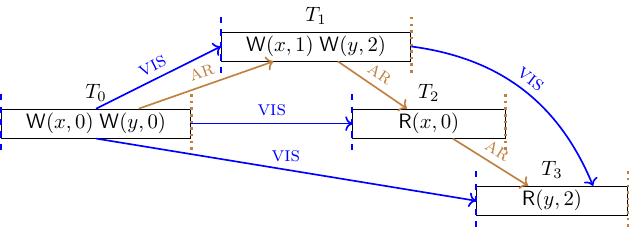}
	\caption{Illustration of both the operational
		and axiomatic semantics of SI.
	  The start and commit timestamps
		are represented by dashed lines and dotted lines, respectively.
		All timestamps are totally ordered from left to right.
	  ({The arrows for $\ar$ implied
		by $\vis$ and transitivity are not shown.})}
	\label{fig:si-op-axiomatic}
\end{figure}

\begin{example}
	Figure~\ref{fig:si-op-axiomatic} shows a valid execution history of SI
	according to Algorithm~\ref{alg:si}
	(ignore the edges for now).
	In this history, the snapshot taken by $T_{2}$
	does not contain the effect of $T_{1}$,
	since $T_{1}.\committime > T_{2}.\starttime$.
	In contrast, the snapshot taken by $T_{3}$
	contains the effect of $T_{1}$,
	thereby reading $2$ from $y$ written by $T_{1}$.
\end{example}
%%%%%%%%%%%%%%%%%%%%%%%%%%%%%%
% si-axiomatic.tex

%%%%%%%%%%%%%%%%%%%%%%%%%%%%%%
\subsection{SI: Axiomatic Semantics} \label{ss:si-axiomatic}

To define the axiomatic semantics of SI,
we first introduce the formal definitions of transactions, histories,
and abstract executions~\cite{Framework:CONCUR2015}.
In this paper, we use $(a, b) \in R$ and $a \rel{R} b$ interchangeably,
where $R \subseteq A \times A$ is a relation over the set $A$.
We write $\domain(R)$ for the domain of $R$.
Given two relations $R$ and $S$, their composition is defined as
$R \comp S = \{ (a, c) \mid \exists b \in A: a \rel{R} b \rel{S} c\}$.
A strict partial order is an irreflexive and transitive relation.
A strict total order is a relation that is a strict partial order and total.

\begin{definition} \label{def:transaction}
  A {\it transaction} is a pair $(\opset, \po)$,
  where $\opset$ is a set of operations
  and $\po$ is
  % and $\po \subseteq \opset \times \opset$ is
	the {\it program order} over $\opset$.
\end{definition}

% For a transaction $T$, we let $T \vdash \writeevent(k, v)$
% if $T$ writes to $k$ and the last value written is $v$,
% and $T \vdash \readevent(k, v)$
% if $T$ reads from $k$ before writing to it
% and $v$ is the value returned by the first such read.
% We also use $\WriteTx_{k} = \set{T \mid T \vdash \writeevent(k, \_)}$.

Clients interact with the store by issuing transactions during \emph{sessions}.
We use a \emph{history} to record the client-visible results of such interactions.

\begin{definition} \label{def:history}
  A {\it history} is a pair $\h = (\T, \SO)$,
  where $\T$ is a set of transactions
  and $\SO$
  % and $\SO \subseteq \T \times \T$
	is the {\it session order} over $\T$.
\end{definition}

We assume that every history contains a special transaction $\nulltxn$
that writes the initial values of all keys~\cite{AnalysingSI:JACM2018, Complexity:OOPSLA2019, MonkeyDB:OOPSLA2021}.
This transaction precedes all the other transactions in $\SO$.

To declaratively justify each transaction of a history,
it is necessary to establish two key relations: visibility and arbitration.
The visibility relation defines
the set of transactions whose effects are visible to each transaction,
while the arbitration relation determines
the execution order of transactions.

\begin{definition} \label{def:ae}
  An {\it abstract execution} is a tuple
  $\ae = (\T, \so, \vis, \ar)$, where
  $(\T, \so)$ is a history,
  {\it visibility} $\vis \subseteq \T \times \T$
  is a strict partial order, and
  {\it arbitration} $\ar \subseteq \T \times \T$
  is a strict total order such that $\vis \subseteq \ar$.
\end{definition}

% For $\h = (\T, \so)$,
% we often shorten $(\T, \so, \vis, \ar)$ to $(\h, \vis, \ar)$.
%%%%%%%%%%%%%%%%%%%%
\begin{definition}[Snapshot Isolation
	(Axiomatic)~\cite{Framework:CONCUR2015, AnalysingSI:JACM2018}]
  \label{def:si-axiomatic}
	A history $\h = (T, \SO)$ satisfies SI if and only if
  there exists an abstract execution $\ae = (T, \SO, \vis, \ar)$
  such that the $\sessionaxiom$, $\intaxiom$, $\extaxiom$,
	$\prefixaxiom$, and $\noconflictaxiom$ axioms (explained below) hold.
\end{definition}

Let $r \triangleq \readevent(k, \_)$ be a read operation of transaction $T$.
If $r$ is the first operation on $k$ in $T$,
then it is called an \emph{external} read operation of $T$;
otherwise, it is an \emph{internal} read operation.

The \emph{internal consistency axiom} \intaxiom{} ensures that,
within a transaction, an internal read from a key returns the same value
as the last write to or read from this key in the transaction.
The \emph{external consistency axiom} \extaxiom{} ensures that
an external read in a transaction $T$ from a key
returns the final value written by the last transaction in $\ar$
among all the transactions that precede $T$ in terms of $\vis$
and write to this key.

The \sessionaxiom{} axiom (i.e., $\SO \subseteq \vis$),
requires previous transactions be visible to
transactions later in the same session.\footnote{
That is, we consider the \emph{strong session} variant
of \si{}~\cite{LazyReplSI:VLDB2006, AnalysingSI:JACM2018}
commonly used in practice.}
The \prefixaxiom{} axiom (i.e., $\ar \comp \vis \subseteq \vis$) ensures that
if the snapshot taken by a transaction $T$ includes a transaction $S$,
than this snapshot also include all transactions that committed before $S$
in terms of $\ar$.
The \noconflictaxiom{} axiom
prevents concurrent transactions from writing on the same key.
That is, for any conflicting transactions $S$ and $T$,
one of $S \rel{\vis} T$ and $T \rel{\vis} S$ must hold.

\begin{example}
  \label{ex:si-axiomatic}

  Figure~\ref{fig:si-op-axiomatic} shows that the history
  is valid under the axiomatic semantics of SI.
  It constructs a possible abstract execution
  by establishing the visibility and arbitration relations
  as shown in the figure.
  For example, $T_{0}$ is visible to $T_{2}$ but $T_{1}$ is not.
  Both $T_{0}$ and $T_{1}$ are visible to $T_{3}$.
  Since $T_{1}$ is after $T_{0}$ in terms of $\ar$,
  $T_{3}$ reads the value 2 of $y$ written by $T_{1}$.
\end{example}
%%%%%%%%%%%%%%%%%%%%%%%%%%%%%%

% which additionally requires a transaction to
% observe all the effects of the preceding transactions
% in the same session.
% \red{Many production databases}, including TiDB~\cite{TiDB},
% Dgraph~\cite{dgraph}, Galera~\cite{maria-galera},
% and CockroachDB~\cite{cockroach}, provide this isolation level.
%%%%%%%%%%%%%%%%%%%%%%%%%%%%%%
% ts-checker.tex

%%%%%%%%%%%%%%%%%%%%%%%%%%%%%%
\section{Timestamp-based SI Checking Algorithms}
\label{section:ts-checker}

We first describe the insights
behind our timestamp-based SI checking algorithms
and then propose both the offline and online checkers.

%%%%%%%%%%%%%%%%%%%%
% si-op-axiomatic.tex

%%%%%%%%%%%%%%%%%%%%%%%%%%%%%%
\subsection{Insights}
\label{ss:insights}

Consider a database history $\H = (\T, \so)$
generated by Algorithm~\ref{alg:si}.
We need to check whether $\H$ satisfies SI
by constructing an abstract execution $\ae = (\T, \so, \vis, \ar)$
with appropriate relations $\vis$ and $\ar$ on $\T$.
The key insight of our approach is to
infer the actual execution order of transactions
to be consistent with their commit timestamps.
% from their commit timestamps.
Formally, we define $\ar$ as follows.
\begin{definition}[Timestamp-based Arbitration]
	\label{def:ar-ts}
	\begin{align*}
		\label{eq:ar-ts}
		\forall T_{1}, T_{2} \in \T.\;
		  T_{1} \rel{\ar} T_{2} \iff T_{1}.\committime < T_{2}.\committime.
	\end{align*}
\end{definition}

On the other hand, from start timestamps
and the inferred execution order,
we can determine the snapshot of each transaction:
each transaction observes the effects of all transactions
that have committed before it starts.
Formally, we define $\vis$ as follows.

\begin{definition}[Timestamp-based Visibility]
	\label{def:vis-ts}
	\begin{align*}
		% \label{eq:vis-ts}
		\forall T_{1}, T_{2} \in \T.\;
		  T_{1} \rel{\vis} T_{2} \iff
			T_{1}.\committime \le T_{2}.\starttime.
	\end{align*}
\end{definition}

\begin{example}
	\label{ex:si-op-aximatic}

	The $\vis$ and $\ar$ relations of Figure~\ref{fig:si-op-axiomatic}
	are constructed according to Definitions~\ref{def:vis-ts} and~\ref{def:ar-ts}, respectively.
	For example, four transactions in the history are totally ordered
	(in $\ar$) by their commit timestamps.
	On the other hand, since $T_{1}.\committime < T_{3}.\starttime$,
	we have $T_{1} \rel{\vis} T_{3}$.
\end{example}

% \begin{theorem}
% 	\label{thm:vis-po}
% 	$\vis$ is a strict partial order.
% \end{theorem}

% \begin{theorem}
% 	\label{thm:ar-to}
% 	$\ar$ is a strict total order and $\vis \subseteq \ar$.
% \end{theorem}

% \begin{proof}
% 	\label{proof:ar-to}
% 	It is easy to see that $\ar$ is a strict total order.
% 	Let $T_{1}, T_{2} \in \T$.
% 	Suppose that $T_{1} \rel{\vis} T_{2}$.
% 	By Definition~\ref{def:vis-ts},
% 	$T_{1}.\committime \le T_{2}.\starttime$.
% 	Since $T_{2}.\starttime < T_{2}.\committime$,
% 	$T_{1}.\committime < T_{2}.\committime$.
% 	By Definition~\ref{def:ar-ts}, $T_{1} \rel{\ar} T_{2}$.
% \end{proof}

% \begin{theorem}
% 	\label{thm:prefixaxiom}
% 	The \prefixaxiom{} axiom holds, i.e., $\ar\comp\vis \subseteq \vis$.
% \end{theorem}

% \begin{proof}
% 	\label{proof:prefixaxiom}
% 	Consider $T_{1}, T_{2}, T_{3} \in \T$.
% 	Suppose that $T_{1} \rel{\ar} T_{2} \rel{\vis} T_{3}$.
% 	By Definitions~\ref{def:ar-ts} and~\ref{def:vis-ts},
% 	\[
% 		T_{1}.\committime < T_{2}.\committime \le T_{3}.\starttime.
% 	\]
% 	Therefore, by Definition~\ref{def:vis-ts},
% 	$T_{1} \rel{\vis} T_{3}$.
% \end{proof}

Given the $\vis$ and $\ar$ relations,
it is straightforward to show that the \prefixaxiom{} axiom holds.
% , i.e., $\ar\comp\vis \subseteq \vis$.
Therefore, by Definition~\ref{def:si-axiomatic},
it remains to check that the \sessionaxiom, \intaxiom, \extaxiom,
and \noconflictaxiom{} axioms hold.
Since transactions are totally ordered,
our checking algorithm can \emph{simulate}
transaction execution one by one,
following their commit timestamps,
and check the axioms on the fly.
This simulation approach enables highly efficient checking algorithms
that eliminate the need to explore all potential
transaction execution orders within a history,
as required by PolySI~\cite{PolySI:VLDB2023} and Viper~\cite{Viper:EuroSys2023},
or to conduct expensive cycle detection
on extensive dependency graphs of histories,
as required by Elle~\cite{Elle:VLDB2020} and Emme~\cite{Clark:EuroSys2024}.
% Moreover, this simulation is not restricted to
% specific data types or the \uniqueval{} assumption,
% as long as the sequential semantics of the data type is provided.
Moreover, the incremental nature of the simulation
renders our algorithm suitable for online checking.
%%%%%%%%%%%%%%%%%%%%%%%%%%%%%%
% offline.tex

%%%%%%%%%%%%%%%%%%%%%%%%%%%%%%
\subsection{The \tool{} Offline Checking Algorithm}
\label{ss:chronos}

%%%%%%%%%%%%%%%%%%%%%%%%%%%%%%
% chronos.tex

% \small
\begin{algorithm}[t]
\footnotesize
	\caption{\tool: the offline SI checking algorithm}
	\label{alg:chronos}
	\begin{algorithmic}[1]
		\Statex $\lastsnoinsession \in \SID \to \N$:
		  $\sno$ of the last transaction already processed
			\Statex \quad in each session, initially $-1$
		\Statex $\lastctsinsession \in \SID \to \TS$:
		  $\committime$ of the last transaction already
			\Statex \quad processed in each session, initially $\nullts$

		% \hStatex
		\Statex $\frontier \in \Key \to \Val$:
			{the last committed value of each key, initially $\nullval$}
		\Statex $\ongoing \in \Key \to 2^{\TID}$:
		{the set of ongoing transactions on each key,}
			\Statex {\quad initially $\emptyset$}

		% \hStatex
		\Statex $\intval \in \TID \to (\Key \to \Val)$:
			the last read/written value of each key
				\Statex \quad in each transaction, {initially $\nullval$}

		\Statex $\extval \in \TID \to (\Key \to \Val)$:
			the last written value of each key
				\Statex \quad in each transaction, {initially $\nullval$}

		\hStatex
		\Procedure{\checksi}{$\H$} \Comment{$\H$ contains
		  the initial transaction $\nulltxn$}
			\label{line:proc-checksi}
			% \State add the initial transaction $\nulltxn$ to $\H$
			%    \label{line:checksi-init-so}

			\State sort $\TS$ in ascending order
				\label{line:checksi-sort-ts}

			\For{$\tsvar \in \TS$}
				\label{line:checksi-foreach-ts}
				\State $T \gets \text{ the transaction that owns the timestamp } \tsvar$
					\label{line:checksi-t}
				\State $\sid \gets T.\sid$ \qquad $\tid \gets T.\tid$
				  \qquad $T.\wkey \gets \emptyset$
				  \label{line:checksi-sid-tid}
					\label{line:checksi-t-wkey}

				\hStatex
				\If{$\tsvar = T.\starttime$} \Comment{start event of $T$}
				  \label{line:checksi-ts-startts}
					\If{{$T.\sno \neq \lastsnoinsession[\sid] + 1
					       \lor T.\starttime < \lastctsinsession[\sid]$}}
						\label{line:checksi-sessionaxiom-begin}
						\State {\it report a violation of \sessionaxiom}
					\EndIf
					\State $\lastsnoinsession[\sid] \gets T.\sno$
					\State $\lastctsinsession[\sid] \gets T.\committime$
						\label{line:checksi-sessionaxiom-end}

					\hStatex
					\For{$(\opvar = \_(k, v)) \in T.\ops$}
						\Comment{in program order}
						\If{$\opvar = \readevent(k, v)$}
							\label{line:checksi-rop}
							\label{line:checksi-intext-begin}
							\If{$\intval[\tid][k] = \nullval$}
								\Comment{external read}
								\label{line:checksi-ext-read}
								\If{{$v \neq \frontier[k]$}}
									\State {\it report a violation of \extaxiom}
									\label{line:checksi-extaxiom-violation}
								\EndIf
							\ElsIf {$\intval[\tid][k] \neq v$}
								\Comment{internal read}
								\label{line:checksi-int-read}
								\State {\it report a violation of \intaxiom}
								\label{line:checksi-intaxiom-violation}
							\EndIf
							\label{line:checksi-intext-end}
						\Else	\Comment{$\opvar = \writeevent(k, v)$}
							\label{line:checksi-wop}
							\label{line:checksi-wop-begin}
							\State $T.\wkey \gets T.\wkey \cup \set{k}$
								\label{line:checksi-wkey-k}
							\State $\extval[\tid][k] \gets v$
								\label{line:checksi-ext-wkey}
							\State {{{$\ongoing[k] \gets \ongoing[k] \cup \set{\tid}$}}}
								\label{line:checksi-ongoing-t-cup}
								\label{line:checksi-wop-end}
						\EndIf
						\State $\intval[\tid][k] \gets v$
							\label{line:checksi-intval}
					\EndFor
				\Else \Comment{commit event of $T$}
				  \label{line:checksi-ts-committs}
					\If{{$T.\starttime > T.\committime$}}
					  \label{line:checksi-startts-committs-begin}
						\Comment{{check Eq. (\ref{eq:si-startts-committs})}}
						\State {\it report an error}
							\label{line:checksi-startts-committs-end}
					\EndIf
					\For{$k \in T.\wkey$}
					  \label{line:checksi-each-k}
						\State $\ongoing[k] \gets \ongoing[k] \setminus \set{\tid}$
						  \Comment{except $T$ itself}
							\label{line:checksi-ongoing-t-setminus}
						\label{line:checksi-noconflictaxiom-begin}
							\If{$\ongoing[k] \neq \emptyset$}
								\label{line:checksi-noconflictaxiom-conflict}
								\State {\it report a violation of \noconflictaxiom}
								\label{line:checksi-noconflictaxiom-violation}
								\label{line:checksi-noconflictaxiom-end}
							\EndIf
							\State $\frontier[k] \gets \extval[\tid][k]$
								\label{line:checksi-frontier-k}
					\EndFor
					% \State $\gcstyle{T.\wkey \gets \emptyset}$
					%   \Comment{gc $\wkey$}
					% 	\label{line:checksi-gc-wkey}
						\label{line:checksi-gc-begin}
					\State $\gcstyle{\intval[\tid] \gets \emptyset}$
						\label{line:checksi-gc-intval}
						\Comment{gc $\intval$}
					\State $\gcstyle{\extval[\tid] \gets \emptyset}$
						\label{line:checksi-gc-extval}
						\Comment{gc $\extval$}
					\State $\gcstyle{\T \gets \T \setminus \set{T}}$
					  \label{line:checksi-gc-txn}
						\Comment{gc $T$}
						\label{line:checksi-gc-end}
				\EndIf
			\EndFor
		\EndProcedure
	\end{algorithmic}
\end{algorithm}
\normalsize

%%%%%%%%%%%%%%%%%%%%%%%%%
\subsubsection{Input}
\label{sss:chronos-transactions}

\tool{} takes a history $\H$ as input and
checks whether it satisfies SI.
Each transaction $T$ in the history is associated with the following data:
\begin{itemize}
	\item $T.\tid$: the unique ID of $T$.
		We use $\TID$ to denote the set of all transaction IDs in $\H$;
	\item $T.\sid$: the unique ID of the session to which $T$ belongs.
		We use $\SID$ to denote the set of all session IDs in $\H$;
	\item $T.\sno$: the sequence number of $T$ in its session;
		% starting from $0$;
	\item $T.\ops$: the list of operations in $T$; and
	\item $T.\starttime$, $T.\committime$:
		the start and commit timestamp of $T$, respectively.
		We use $\TS$ to denote the set of all timestamps in $\H$
		and denote the minimum timestamp in $\TS$ by $\nullts$.
\end{itemize}
We use $T.\wkey$ to record the set of keys written by $T$.
We design \tool{} with key-value histories in mind,
but it is also easily adaptable to support other data types
such as lists.
% We also use $\opvar.\keyvar$ to denote the key of an operation $\opvar$.
%%%%%%%%%%%%%%%%%%%%%%%%%
\subsubsection{Algorithm}
\label{sss:chronos-algorithm}

\tool{} simulates the execution of a database
assuming that the start and commit events of transactions in $\H$
are executed in the order of their timestamps,
while checking the corresponding axioms on the fly.
To do this,
% \tool{} first adds the initial transaction $\nulltxn$ to $\H$,
% which precedes all the transactions in $\H$ in SO
% (\code{\ref{alg:chronos}}{\ref{line:checksi-init-so}}).
% Then
\tool{} sorts all the timestamps in $\TS$ in ascending order
(\code{\ref{alg:chronos}}{\ref{line:checksi-sort-ts}}).
By Definition~\ref{def:ar-ts}, we obtain the total order $\ar$ between transactions.
\tool{} will traverse $\TS$ in this order
and process each of the start events and commit events of transactions one by one
(\code{\ref{alg:chronos}}{\ref{line:checksi-foreach-ts}}).
During this process, \tool{} maintains two maps:
{$\frontier$} maps each key to the last (in $\ar$) committed value,
and {$\ongoing$} maps each key to the set of ongoing transactions
that write this key.

Let $\tsvar$ be the current timestamp being processed
and $T$ the transaction that owns the timestamp $\tsvar$.
If $\tsvar$ is the start timestamp of $T$,
\tool{} checks whether $T$ violates any of
the \sessionaxiom, \intaxiom, and \extaxiom{} axioms
(\code{\ref{alg:chronos}}{\ref{line:checksi-ts-startts}}).
Otherwise, it checks whether $T$ violates the \noconflictaxiom{} axiom
(\code{\ref{alg:chronos}}{\ref{line:checksi-ts-committs}}).
% Finally, \tool{} periodically triggers a garbage collection (GC).
\tool{} will identify \emph{all} violations of these axioms in a history,
instead of terminating immediately upon encountering the first one.
%%%%%%%%%%%%%%%%%%%%
\subsubsection*{\sessionaxiom~(\codes{\ref{alg:chronos}}{\ref{line:checksi-sessionaxiom-begin}}{\ref{line:checksi-sessionaxiom-end}})}

\tool{} uses \lastsnoinsession{} and \lastctsinsession{}
to maintain the $\sno$ and $\committime$ of the last transaction
already processed in each session, respectively.
It checks whether the current transaction $T$
follows immediately after the last transaction already processed in its session
and starts after this last transaction commits.
If this fails, a violation of \sessionaxiom{} is found.
%%%%%%%%%%%%%%%%%%%%
\subsubsection*{\intaxiom{} and \extaxiom~(\codes{\ref{alg:chronos}}{\ref{line:checksi-intext-begin}}{\ref{line:checksi-intext-end}})}

Let $\tid \triangleq T.\tid$.
\tool{} uses $\intval[\tid][k]$ to track the last value
read or written by $T$ on key $k$
(see \code{\ref{alg:chronos}}{\ref{line:checksi-intval}}).
Let $\opvar = \readevent(k, v)$ be the read operation of $T$
being checked~(\code{\ref{alg:chronos}}{\ref{line:checksi-rop}}).
If $\intval[\tid][k]$ is not the initial artificial value $\nullval$,
then $\opvar$ is an internal read.
\tool{} then checks if $\intval[\tid][k] = v$.
If this fails, a violation of \intaxiom{} is found
(\code{\ref{alg:chronos}}{\ref{line:checksi-intaxiom-violation}}).

If $\intval[\tid][k] = \nullval$,
then $\opvar$ is an external read~(\code{\ref{alg:chronos}}{\ref{line:checksi-ext-read}}).
By \extaxiom{}, $\opvar$ should observe the last committed value of $k$,
i.e., $\frontier[k]$.
Otherwise, a violation of \extaxiom{} is found
(\code{\ref{alg:chronos}}{\ref{line:checksi-extaxiom-violation}}).

For a key $k$, \tool{} updates $\frontier[k]$
whenever a transaction that writes to $k$ commits.
Since a transaction $T$ may write to the same key $k$ multiple times,
\tool{} uses $\extval[\tid][k]$ to track the last value written by $T$ on $k$
(see \code{\ref{alg:chronos}}{\ref{line:checksi-ext-wkey}}).
Therefore, when $T$ commits,
\tool{} updates $\frontier[k]$ to $\extval[\tid][k]$
(\code{\ref{alg:chronos}}{\ref{line:checksi-frontier-k}}).
%%%%%%%%%%%%%%%%%%%%
\subsubsection*{\noconflictaxiom~(\codes{\ref{alg:chronos}}{\ref{line:checksi-noconflictaxiom-begin}}{\ref{line:checksi-noconflictaxiom-end}})}

Now suppose that $\tsvar$ is the commit timestamp of $T$
(\code{\ref{alg:chronos}}{\ref{line:checksi-ts-committs}}).
For each key $k \in T.\wkey$ written by $T$
(tracked at \code{\ref{alg:chronos}}{\ref{line:checksi-wkey-k}}),
\tool{} checks whether $T$ conflicts with any ongoing transactions on key $k$
by testing the emptiness of $\ongoing[k]$
(tracked at \code{\ref{alg:chronos}}{\ref{line:checksi-ongoing-t-cup}}).
If $\ongoing[k]$ (except $T$ itself) is non-empty,
a violation of \noconflictaxiom{} is found~(\code{\ref{alg:chronos}}{\ref{line:checksi-noconflictaxiom-violation}}).
% Additionally, \tool{} updates $\ongoing$
% (\code{\ref{alg:chronos}}{\ref{line:checksi-ongoing-t-setminus}})
% and $\frontier$ (\code{\ref{alg:chronos}}{\ref{line:checksi-frontier-k}}) accordingly.
%%%%%%%%%%%%%%%%%%%%
\subsubsection*{\gc~(\codes{\ref{alg:chronos}}{\ref{line:checksi-gc-begin}}{\ref{line:checksi-gc-end}})}

To save memory, \tool{} periodically discard obsolete information
from $\intval$ and $\extval$ and remove old transactions from $\T$.
First, it is safe to discard $\intval[\tid]$
once the commit event of transaction $T$ has been processed
(\code{\ref{alg:chronos}}{\ref{line:checksi-gc-intval}}),
since the information in $\intval$ will never be used
by any other transaction.
Second, for a transaction $T$,
all of its writes have been recorded in $\frontier$,
$\extval[\tid]$ becomes redundant and can be discarded
(\code{\ref{alg:chronos}}{\ref{line:checksi-gc-extval}}).
Furthermore, $T.\sno$ and $T.\cts$ are recorded in \lastsnoinsession{}
and \lastctsinsession{}, respectively,
and thus $T$ is no longer needed in $\T$
(\code{\ref{alg:chronos}}{\ref{line:checksi-gc-txn}}).
%%%%%%%%%%%%%%%%%%%%
\begin{example}
	\label{ex:chronos}
	Consider Figure~\ref{fig:online-challenges}'s history
	which consists of five transactions,
	namely $T_{1}, T_{2}, \ldots$, and $T_{5}$.
	\tool{} processes the start and commit events of the transactions
	in the ascending order of their timestamps,
	namely \startstyle{\circleone}, \commitstyle{\circletwo}, $\ldots$,
	and \commitstyle{\circleten}.

	On the start event of $T_{3}$ with timestamp \startstyle{\circlesix},
	the external read operation $\readevent(x, 2)$ can be justified
	by the last committed transaction $T_{2}$ that writes $2$ to $x$.
	This is captured by the current $\frontier[x] = \set{T_{2}}$.
	Moreover, since $T_{3}$ updates key $y$,
	\tool{} adds $T_{3}$ to $\ongoing[y]$,
	yielding $\ongoing[y] = \set{T_{3}}$.
	Now comes the commit event of $T_{5}$ with timestamp \commitstyle{\circleseven}.
	\tool{} finds that $T_{5}$ conflicts with $T_{3}$ on $y$
	by checking the emptiness of $\ongoing[y] \setminus \set{T_{5}}$,
	reporting a violation of \noconflictaxiom{}.
	In addition, since $T_{5}$ updates key $y$,
	\tool{} updates $\frontier[y]$ to $\set{T_{5}}$.
	This justifies the external read operation $\readevent(y, 1)$
	of $T_{4}$ when \tool{} examines the start event of $T_{4}$
	with timestamp \startstyle{\circleeight}.
	Note that upon the commit event of $T_{3}$
	with timestamp \commitstyle{\circlenine},
	\tool{} does \emph{not} report a violation of \noconflictaxiom{}
	for $T_{3}$ with $T_{5}$ on $y$
	because this violation has already been reported
	on the commit event of $T_{5}$ and
	now $T_{5} \notin \ongoing[y]$
	at~\code{\ref{alg:chronos}}{\ref{line:checksi-ongoing-t-setminus}}.

	% \tool{} continues checking
	% as if the violation of \noconflictaxiom{} does not happen.
	% In this example, \tool{} will
	% and updates $\frontier[y] = \set{T_{3}}$
	% at timestamp \commitstyle{\circlenine}.
	% Suppose that \gc{} is triggered at this point.
	% As indicated by $\frontier[x] = \set{T_{2}}$
	% and $\frontier[y] = \set{T_{3}}$,
	% $T_{1}$ and $T_{5}$ have been superseded by $T_{2}$
	% and $T_{3}$ on $x$ and $y$, respectively
	% (see \code{\ref{alg:chronos}}{\ref{line:gc-not-at-frontier}}).
	% Therefore, it is safe for \tool{} to
	% discard $\extval[T_{1}]$ and $\extval[T_{5}]$.
\end{example}
%%%%%%%%%%%%%%%%%%%%%%%%%
\subsubsection{Complexity Analysis}
\label{sss:chronos-complexity}

Suppose that the history $\H$ consists of
$N$ transactions and $M$ operations ($N \le M$).
We assume that the data structures used by \tool{}
are implemented as hash maps (or hash sets)
which offer average constant time performance for
the basic operations like \textsf{put}, \textsf{get},
\textsf{remove}, \textsf{contains}, and \textsf{isEmpty}.

The time complexity of \tool{} comprises
\begin{itemize}
	\item $O(N \log N)$ for sorting on $\TS$
	  (\code{\ref{alg:chronos}}{\ref{line:checksi-sort-ts}});
	\item $O(N) = N \cdot O(1)$ for checking \sessionaxiom{}
	  on the start events of transactions
	  (\codes{\ref{alg:chronos}}{\ref{line:checksi-sessionaxiom-begin}}{\ref{line:checksi-sessionaxiom-end}});
	\item $O(M) = M \cdot O(1)$ for checking \extaxiom{}
	  and \intaxiom{} on each read operation of transactions
	  (\codes{\ref{alg:chronos}}{\ref{line:checksi-intext-begin}}{\ref{line:checksi-intext-end}})
		and for updating data structures
		on each write operation of transactions
		(\codes{\ref{alg:chronos}}{\ref{line:checksi-wop-begin}}{\ref{line:checksi-wop-end}});
	\item $O(N) = N \cdot O(1)$ for checking Eq. (\ref{eq:si-startts-committs})
	  on the commit events of transactions
	  (\codes{\ref{alg:chronos}}{\ref{line:checksi-startts-committs-begin}}{\ref{line:checksi-startts-committs-end}});
	\item $O(M) = O(M) \cdot O(1)$ for checking \noconflictaxiom{}
	  on the commit events of transactions
	  (\codes{\ref{alg:chronos}}{\ref{line:checksi-noconflictaxiom-begin}}{\ref{line:checksi-noconflictaxiom-end}}).
\end{itemize}
Therefore, the time complexity is $O(N \log N + M)$.

The space complexity of \tool{} is primarily determined by
the memory required for storing the history
(the memory usage of the $\frontier$ and $\ongoing$ data structures
is relatively small).
Thanks to its prompt garbage collection mechanism,
the memory usage for storing the history diminishes
as transactions are processed.
%%%%%%%%%%%%%%%%%%%%%%%%%%%%%%
% online.tex

%%%%%%%%%%%%%%%%%%%%%%%%%%%%%%
\subsection{The \onlinetool{} Online Checking Algorithm}
\label{ss:aion}

%%%%%%%%%%%%%%%%%%%%%%%%%
\subsubsection{Input}
\label{sss:input}

In the online settings, the history $\H$ is not known in advance.
Instead, the online checking algorithm \onlinetool{}
receives transactions one by one and checks \si{} incrementally.
It is assumed that the session order is preserved
when transactions are received.

%%%%%%%%%%%%%%%%%%%%%%%%%
\subsubsection{Challenges} \label{sss:online-challenges}

Due to asynchrony, \onlinetool{} \emph{cannot} anticipate that
transactions will be collected in ascending order
based on their start/commit timestamps.
This introduces three main challenges for \onlinetool.

\begin{itemize}
	\item The determination of satisfaction or violation
		of the \extaxiom{} axiom for a transaction
		becomes \emph{unstable} due to asynchrony.
		This means that we cannot assert it immediately
		upon the transaction being collected.
		In contrast, the \intaxiom{} axiom remains unaffected by asynchrony,
		while \noconflictaxiom{} violations will eventually be correctly reported
		as soon as the conflicting transactions are collected.
	\item An incoming transaction $T$ with a smaller start timestamp
		may initiate the \emph{re-checking} of transactions
		that commit after $T$ starts.
		To facilitate efficient rechecking,
		it is essential to extend and maintain the data structures,
		namely \frontier{} and \ongoing{}.
	\item To prevent unlimited memory usage
	  and facilitate long-running checking,
		\onlinetool{} must recycle data structures
		and transactions that are no longer necessary.
		Unfortunately, in the worst-case scenario,
		\onlinetool{} cannot safely recycle
		any data structures and transactions due to asynchrony.
\end{itemize}

% online-challenges.tex

\begin{figure}[t]
	\centering
	\includegraphics[width = 0.32\textwidth]{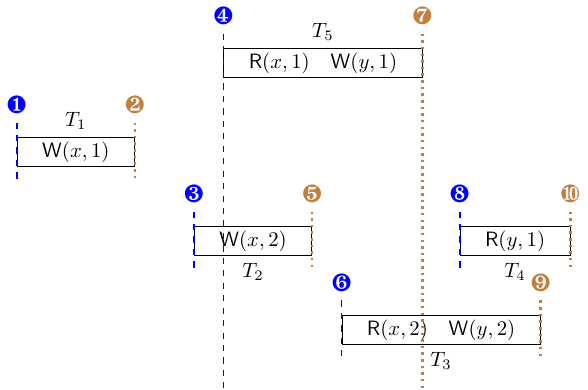}
	\caption{Illustration of both the offline and online checking algorithms, \tool{} and \onlinetool{} in Example~\ref{ex:chronos} and Example~\ref{ex:aion}.}
	\label{fig:online-challenges}
\end{figure}

% \hfwei{adding a transaction entirely within $T_{5}$}}

The following example illustrates the challenges faced by \onlinetool{}
and our solutions to overcome them.
\begin{example}
	\label{ex:aion}
	Consider the history of Figure~\ref{fig:online-challenges}.
	Suppose that the transactions are collected in the order
	$T_{1}$, $T_{2}$, $\ldots$, and $T_{5}$.

	When \onlinetool{} collects the first four transactions,
	it cannot definitively assert that $T_{4}$ violates the \extaxiom{} axiom permanently.
	This uncertainty arises because $T_{5}$,
	from which $T_{4}$ reads the value of $y$,
	may experience delays due to asynchrony.
	To tackle this instability issue,
	\onlinetool{} employs a {waiting period}
	during which delayed transactions are expected to be collected.
	This approach allows for a more accurate determination of
	whether a violation of \extaxiom{} persists
	or is a transient result.

	Now, when transaction $T_{5}$ arrives,
	\onlinetool{} must check the \sessionaxiom, \intaxiom,
	and \extaxiom{} axioms for $T_{5}$
	and \emph{re-check} all transactions that commit after $T_{5}$ starts
	(denoted \startstyle{\circlefour}).
	The re-checking process involves two cases
	based on the axioms to be verified:
		(1) re-check the \noconflictaxiom{} axiom
		  for transactions overlapping with $T_{5}$,
			i.e., $T_{2}$ and $T_{3}$ in this example; and
		(2) re-check the \extaxiom{} axiom
		  for transactions starting after $T_{5}$ commits,
		  i.e., $T_{4}$ in this example.
	To facilitate this, the data structures \frontier{} and \ongoing{}
	should be \emph{versioned} by timestamps and support timestamp-based search,
	returning the latest version \emph{before} a given timestamp.

	For instance, to determine the set of transactions
	overlapping with $T_{5}$ (for checking the \noconflictaxiom{} axiom),
	\onlinetool{} queries the versioned data structure \ongoing{}
	using the timestamp $T_{5}.\committime$ \commitstyle{\circleseven} (and key $y$).
	This query returns the set of ongoing transactions
	updated at timestamp $T_{3}.\starttime$ \startstyle{\circlesix},
	namely $\set{T_{3}.\tid}$.
	Consequently, \onlinetool{} identifies a conflict between $T_{5}$ and $T_{3}$
	and reports a violation of \noconflictaxiom.

	Similarly, to justify the read operation $\readevent(y, 1)$ of $T_{4}$
	(for re-checking the \extaxiom{} axiom),
	\onlinetool{} queries the versioned data structure \frontier{}
	using the timestamp $T_{4}.\starttime$ \startstyle{\circleeight} (and key $y$).
	This query returns the frontier updated at timestamp $T_{5}.\committime$
	\commitstyle{\circleseven}, namely $[y \mapsto 1]$.
	Thus, $T_{4}$ is re-justified by $T_{5}$,
	clearing the false alarm on the violation of \extaxiom{} axiom for $T_{5}$.
	Be cautious that this result is also temporary
	and subject to change due to asynchrony.

	For violations of the \intaxiom{}, \extaxiom{},
	and \noconflictaxiom{} axiom,
	\onlinetool{} reports the violation and continues checking
	as if the violation did not occur.
	% imitating the behavior of a malfunctioning database system.
\end{example}
%%%%%%%%%%%%%%%%%%%%%%%%%
\subsubsection{Algorithm} \label{sss:online-algorithm}

% aion.tex

\begin{algorithm*}[t]
        \footnotesize	
	\caption{\onlinetool: the online SI checking algorithm}
	\label{alg:aion}
	\begin{varwidth}[t]{0.50\textwidth}
	\begin{algorithmic}[1]
		\Statex $\lastsnoinsession$, $\lastctsinsession$,
		  $\intval$, and $\extval$ are the same as in \tool{}
		% \Statex $\lastsnoinsession \in \SID \to \N$:
		%   $\sno$ of the last transaction already processed
		% 	\Statex \quad in each session, initially $-1$
		% \Statex $\lastctsinsession \in \SID \to \TS$:
		%   $\committime$ of the last transaction already
		% 	\Statex \quad processed in each session, initially $\nullts$

		% \hStatex
		% \Statex $\frontiertsstyle{\frontierts \in \TS \to (\Key \to \Val)}$:
		% 	$\frontier$ versioned by timestamps
		% \Statex $\ongoingtsstyle{\ongoingts \in \TS \to (\Key \to 2^{\TID})}$:
		%   $\ongoing$ versioned by timestamps
		% \Statex $\frontiertsstyle{\frontierts \in \TS \to (\Key \to \Val)}$:
		% 	the last (in $\ar$) committed value
		% 	\Statex \quad for each key at a timestamp,
		% 	initially $\nullval$ % $[\lambda k.\; k \mapsto \nulltxn]$
		% \Statex $\ongoingtsstyle{\ongoingts \in \TS \to (\Key \to 2^{\TID})}$:
		% 	{the set of ongoing transactions}
		% 	\Statex {\quad on each key at a timestamp,
		% 	initially $\emptyset$} % $[\lambda k.\; k \mapsto \emptyset]$

		% \hStatex
		% \Statex $\intval \in \TID \to (\Key \to \Val)$:
		% the last read/written value of each key
		% 	\Statex \quad in each transaction, {initially $\nullval$}
		% \Statex $\extval \in \TID \to (\Key \to \Val)$:
		% the last written value of each key
		% 	\Statex \quad in each transaction, {initially $\nullval$}

		\hStatex
		\Procedure{\onlinechecksi}{$T$}
		  \Comment{upon receiving a new transaction $T$}
			\label{line:proc-onlinechecksi}
			% \While{receive a new transaction $T$}
				\State \clockstyle{} $T.\extaxiom \gets \top$
				  \Comment{$T.\extaxiom$: whether $T$ satisfies \extaxiom, initially \textsf{true}}
				  \label{line:onlinechecksi-extaxiom-top}
			  \State {\clockstyle{} set a timer for re-checking \extaxiom{} for $T$}
				  \label{line:onlinechecksi-set-timer}
				\If{{$T.\starttime > T.\committime$}}
					\Comment{{check Eq. (\ref{eq:si-startts-committs})}}
					\State {\it report an error}
				\EndIf

				\State $\sid \gets T.\sid$ \qquad $\tid \gets T.\tid$
				  \qquad $T.\wkey \gets \emptyset$
				  \label{line:onlinechecksi-sid-tid}
					\label{line:onlinechecksi-t-wkey}
				\hStatex
				\Statex \Comment{\emph{\ding{172} check \sessionaxiom, \intaxiom, and \extaxiom{}
				  for $T$, similarly to \tool\phantom{\qquad\;\;}}}
					\label{line:onlinechecksi-step1-begin}
				\If{{$T.\sno \neq \lastsnoinsession[\sid] + 1
								\lor T.\starttime < \lastctsinsession[\sid]$}}
					\label{line:onlinechecksi-sessionaxiom-begin}
					\State {\it report a violation of \sessionaxiom}
				\EndIf
				\State $\lastsnoinsession[\sid] \gets T.\sno$
				\State $\lastctsinsession[\sid] \gets T.\committime$
					\label{line:onlinechecksi-sessionaxiom-end}

				\hStatex
				\For{$(\opvar = \_(k, v)) \in T.\ops$}
					\If{$\opvar = \readevent(k, v)$}
						\label{line:onlinechecksi-rop}
						\label{line:onlinechecksi-intext-begin}
						\If{$\intval[\tid][k] = \nullval$}
							\Comment{external read}
							\label{line:onlinechecksi-ext-read}
							\If{$v \neq \frontiertsstyle{\frontierts[\widehat{T.\starttime}][k]}$ \loadstyle}
							  \label{line:onlinechecksi-extaxiom-frontierts-starttime}
								\State \clockstyle{} $T.\extaxiom \gets \bot$
								\label{line:onlinechecksi-extaxiom-violation}
							\EndIf
						\ElsIf {$\intval[\tid][k] \neq v$}
							\Comment{internal read}
							\label{line:onlinechecksi-int-read}
							\State {\it report a violation of \intaxiom}
							\label{line:onlinechecksi-intaxiom-violation}
						\EndIf
						\label{line:onlinechecksi-intext-end}
					\Else	\Comment{$\opvar = \writeevent(k, v)$}
						\label{line:onlinechecksi-wop}
						\State $T.\wkey \gets T.\wkey \cup \set{k}$
							\label{line:onlinechecksi-wkey-k}
						\State $\extval[\tid][k] \gets v$
							\label{line:onlinechecksi-ext-wkey}
					\EndIf
					\State $\intval[\tid][k] \gets v$
						\label{line:onlinechecksi-intval}
				\EndFor

				\State $\cts \gets T.\committime$
				  \label{line:onlinechecksi-cts}
				\State $\frontiertsstyle{\frontierts[\cts] \gets \frontierts[\beforects]}$ \loadstyle
				  \label{line:onlinechecksi-frontier-cts}
				\State {$\frontiertsstyle{\frontierts[\cts][k] \gets \extval[\tid][k]}$
					for $k \in T.\wkey$}
					\label{line:onlinechecksi-frontier-cts-wkey}
				\State {${\intval[\tid][k] \gets \nullval}$ for $k \in \domain(\intval[\tid])$}
					% \label{line:onlinechecksi-gc-intval}
					\Comment{reset}
					\label{line:onlinechecksi-step1-end}
				\hStatex
				\Statex \Comment{\emph{\ding{173} re-check \noconflictaxiom{} for transactions
				  overlapping with $T$} \phantom{\qquad}}
				\For{$\tsvar \in \TS$ such that $T.\starttime \le \tsvar \le T.\committime$}
				  \label{line:onlinechecksi-noconflictaxiom-begin}
					\label{line:onlinechecksi-step2-begin}
				  % \Statex \Comment{in increasing order}
					\State $T' \gets \text{ the transaction that owns the timestamp } \tsvar$ \loadstyle
						\label{line:onlinechecksi-tprime}
					\If{$\tsvar = T'.\starttime$}
					  \For{$k \in T'.\wkey$}
							\State $\ongoingtsstyle{\ongoingts[\tsvar][k] \gets
								\ongoingts[\beforets][k] \cup \set{T'.\tid}}$ \loadstyle
								\label{line:onlinechecksi-ongoingts-tprime-cup}
						\EndFor
					\Else \Comment{$\tsvar = T'.\committime$}
					  \For{$k \in T'.\wkey$}
							\State $\ongoingtsstyle{\ongoingts[\tsvar][k] \gets
								\ongoingts[\beforets][k] \setminus \set{T'.\tid}}$ \loadstyle
							\label{line:onlinechecksi-ongoingts-t-setminus}
							\If{$\ongoingtsstyle{\ongoingts[\ts][k] \neq \emptyset}$}
								\label{line:onlinechecksi-noconflictaxiom-conflict}
								\State {\it report a violation of \noconflictaxiom}
								\label{line:onlinechecksi-noconflictaxiom-violation}
								\label{line:onlinechecksi-noconflictaxiom-end}
								\label{line:onlinechecksi-step2-end}
							\EndIf
						\EndFor
					\EndIf
				\EndFor
				\algstore{onlinechecksi}
			\end{algorithmic}
			\end{varwidth}\quad
			\begin{varwidth}[t]{0.48\textwidth}
			\begin{algorithmic}[1]
				\algrestore{onlinechecksi}
				\Statex $\frontiertsstyle{\frontierts \in \TS \to (\Key \to \Val)}$:
					$\frontier$ versioned by timestamps
				\Statex $\ongoingtsstyle{\ongoingts \in \TS \to (\Key \to 2^{\TID})}$:
					$\ongoing$ versioned by timestamps

				\hStatex
				\Statex \Comment{\emph{\ding{174} re-check \extaxiom{} for transactions
				  starting after $T$ commits\phantom{\qquad\quad\;}}}
				\State \recheckextoptstyle{$\keysvar \gets T.\wkey$}
				  \label{line:onlinechecksi-extaxiom-keys}
				\For{$\tsvar \in \TS$ such that $\tsvar > T.\committime$}
					\label{line:onlinechecksi-step3-begin}
					\State $T' \gets \text{ the transaction that owns the timestamp } \tsvar$ \loadstyle
					\If{$\tsvar = T'.\starttime$}
						\If{\Call{Timeout}{$T'$} has been called}
						  \State \clockstyle{} {\bf continue}
						\EndIf
						\For{$(\opvar = \_(k, v)) \in T'.\ops$}
							\If{$\opvar = \readevent(k, v) \land \intval[T'.\tid][k] = \nullval \land \recheckextoptstyle{k \in \keysvar}$}
							    \label{line:onlinechecksi-extaxiom-ext-read}
									\If{$\extval[\tid][k] \neq v$} % \If{$\extval[T''][k] \neq v$}
										\State \clockstyle{} $T'.\extaxiom \gets \bot$
										  \label{line:onlinechecksi-extaxiom-recheck-bot}
									\Else
										\State \clockstyle{} $T'.\extaxiom \gets \top$
										\label{line:onlinechecksi-extaxiom-recheck-top}
								\EndIf
							\EndIf
							\If{$\recheckextoptstyle{k \in \keysvar}$}
								\State $\recheckextoptstyle{\intval[T'.\tid][k] \gets v}$
							\EndIf
						\EndFor
					\Else \Comment{$\tsvar = T'.\committime$}
						% \State {$\gcstyle{\intval[T'.\tid] \gets \emptyset}$}
						% 	\Comment{gc $\intval$ of $T'$}
						% \State {${\intval[T'.\tid][k] \gets \nullval}$
						%   for $k \in \domain(\intval[T'.\tid])$}
						% 	\Comment{reset $\intval$ of $T'$}
						\For{$k \in \domain(\intval[T'.\tid])$}
							\State {${\intval[T'.\tid][k] \gets \nullval}$}
								\Comment{reset $\intval$ of $T'$}
						\EndFor
						\State \recheckextoptstyle{$\keysvar \gets \keysvar \setminus T'.\wkey$}
						  \label{line:onlinechecksi-extaxiom-keys-setminus}
						\If{\recheckextoptstyle{$\keysvar = \emptyset$}}
							\label{line:onlinechecksi-extaxiom-keys-empty}
						  \State \recheckextoptstyle{\bf break}
							\label{line:onlinechecksi-extaxiom-break}
						\EndIf
						\For{$k \in \keysvar$}
							\State {$\frontiertsstyle{\frontierts[\tsvar][k] \gets \extval[\tid][k]}$}
								\label{line:onlinechecksi-extaxiom-frontierts}
								\label{line:onlinechecksi-step3-end}
						\EndFor
						\State {$\gcstyle{\extval[\tid] \gets \emptyset}$}
							\Comment{gc $\extval$ of $T$}
					\EndIf
				\EndFor
			% \EndWhile
		\EndProcedure

		\Statex
		\Procedure{\textsc{Timeout}}{$T$}
		  \Comment{\clockstyle{} runs when timeout for $T$ expires}
			\label{proc:timeout}
			  \If{$T.\extaxiom = \bot$}
					\State {\it report a violation of \extaxiom}
					\label{line:timeout-extaxiom-violation}
				\EndIf
		\EndProcedure

		\Statex
		\hStatex
		\Comment{gc $\frontierts$, $\ongoingts$,
			and transactions below timestamp $\tsvar$ \phantom{}}
		\Procedure{\textsc{GarbageCollect}}{$\tsvar$}
			\Comment{runs periodically}
			\label{line:online-proc-gc}
			\For{$\tsvar' \le \tsvar$}
				\label{line:online-proc-gc-begin}
				\State $\diskstyle{\frontierts[\tsvar'] \gets \emptyset} $ \savestyle
					\label{line:online-gc-frontierts}
				\State $\diskstyle{\ongoingts[\tsvar'] \gets \emptyset}$ \savestyle
					\label{line:online-gc-ongoingts}
			\EndFor
			\State {$\diskstyle{\T \gets \T \setminus
			  \set{T \in \T \mid T.\committime \le \tsvar}}$} \savestyle
			\label{line:online-gc-T}
			\label{line:online-proc-gc-end}
		\EndProcedure
		% \Statex
		% \Procedure{\textsc{GarbageCollect}}{}
		% 	\Comment{runs periodically}
		% 	\label{line:online-proc-gc}
		% 	\label{line:online-proc-gc-begin}
		% 	\For{$T \in \domain(\extval)$}
		% 		\State $\sts \gets T.\starttime$
		% 		\State $\cts \gets T.\committime$
		% 		\If{$\bigwedge\limits_{k \in \domain(\extval[T])} \!\!\!\!\!\!\!\!
		% 		  \cts < \frontierts[k].\committime$}
		% 			  \label{line:online-gc-not-at-frontierts}
		% 			\State {$\gcstyle{\extval[T] \gets \emptyset}$}
		% 				\Comment{gc $\extval$}
		% 			  \label{line:online-gc-extval}
		% 			\State $\diskstyle{\frontierts[\sts] \gets \emptyset}$
		% 			  \Comment{transfer $\frontierts[\sts]$}
		% 				\label{line:online-gc-frontierts}
		% 			\State $\diskstyle{\ongoingts[\sts] \gets \emptyset}$
		% 			  \Comment{transfer $\ongoingts[\sts]$}
		% 				\label{line:online-gc-ongoingts-sts}
		% 			\State $\diskstyle{\ongoingts[\cts] \gets \emptyset}$
		% 			  \Comment{transfer $\ongoingts[\cts]$}
		% 				\label{line:online-gc-ongoingts-cts}
		% 			\If{$T \neq \lastinsession[T.\sid]$}
		% 				\State {$\diskstyle{\T \gets \T \setminus \set{T}}$}
		% 				\Comment{transfer $T$}
		% 			  \label{line:online-gc-T}
		% 			\EndIf
		% 		\EndIf
		% 	\EndFor
		% 	\label{line:online-proc-gc-end}
		% \EndProcedure
	\end{algorithmic}
	\end{varwidth}
\end{algorithm*}
\normalsize

Algorithm~\ref{alg:aion} provides the pseudocode for \onlinetool.
% the online SI checking algorithm \onlinetool.
\onlinetool{} extends the data structures \frontier{} and \ongoing{}
used in \tool{} to \frontierts{} and \ongoingts{}, respectively,
which are versioned by timestamps.
When given a timestamp $\tsvar$,
we use $\frontierts[\beforets]$ and $\ongoingts[\beforets]$
to denote the latest version of \frontierts{} and \ongoingts{}
\emph{before} $\tsvar$, respectively.

Upon receiving a new transaction $T$,
\onlinetool{} proceeds in three steps:
it first checks the \sessionaxiom, \intaxiom, and \extaxiom{} axioms
for $T$, similarly to \tool{}
(\codes{\ref{alg:aion}}{\ref{line:onlinechecksi-step1-begin}}{\ref{line:onlinechecksi-step1-end}}),
then it re-checks the \noconflictaxiom{} axiom for transactions
overlapping with $T$
(\codes{\ref{alg:aion}}{\ref{line:onlinechecksi-step2-begin}}{\ref{line:onlinechecksi-step2-end}}),
and finally, it re-checks the \extaxiom{} axiom for transactions
starting after $T$ commits
(\codes{\ref{alg:aion}}{\ref{line:onlinechecksi-step3-begin}}{\ref{line:onlinechecksi-step3-end}}).

\subsubsection*{Step \ding{172}~(\codes{\ref{alg:aion}}{\ref{line:onlinechecksi-step1-begin}}{\ref{line:onlinechecksi-step1-end}})}
The checking for $T$ is similar to that in \tool,
with two differences.
On one hand, to check the \extaxiom{} axiom,
\onlinetool{} needs to obtain the latest version of \frontierts{}
before $T.\starttime$, denoted $\frontierts[\widehat{T.\starttime}]$
(\code{\ref{alg:aion}}{\ref{line:onlinechecksi-extaxiom-frontierts-starttime}}).
Moreover, to update \frontierts{} at timestamp $\cts \gets T.\committime$,
it first fetches the latest version of \frontierts{} before $\cts$
(\code{\ref{alg:aion}}{\ref{line:onlinechecksi-frontier-cts}})
and then updates the components for each key written by $T$
(\code{\ref{alg:aion}}{\ref{line:onlinechecksi-frontier-cts-wkey}}).
On the other hand, if $T$ violates the \extaxiom{} axiom,
\onlinetool{} does not report the violation immediately
as it is subject to change due to asynchrony.
Instead, it records the temporary violation in variable $T.\extaxiom$
(\code{\ref{alg:aion}}{\ref{line:onlinechecksi-extaxiom-violation}})
and waits for a timeout (\code{\ref{alg:aion}}{\ref{proc:timeout}})
set at the beginning (\code{\ref{alg:aion}}{\ref{line:onlinechecksi-set-timer}}).

\subsubsection*{Step \ding{173}~(\codes{\ref{alg:aion}}{\ref{line:onlinechecksi-step2-begin}}{\ref{line:onlinechecksi-step2-end}})}
To re-check the \noconflictaxiom{} axiom for transactions overlapping with $T$,
\onlinetool{} iterates through the timestamps $\tsvar$
ranging from $T.\starttime$ to $T.\committime$
(both \emph{inclusive}) in ascending order
(\code{\ref{alg:aion}}{\ref{line:onlinechecksi-noconflictaxiom-begin}}).
If $\tsvar$ corresponds to the start timestamp of a transaction $T'$
(which could be $T$),
\onlinetool{} updates $\ongoingts[\tsvar]$ for each key written by $T'$
(\code{\ref{alg:aion}}{\ref{line:onlinechecksi-ongoingts-tprime-cup}})
to include $T'.\tid$.
If $\tsvar$ corresponds to the commit timestamp of transaction $T'$,
\onlinetool{} reports that $T'$ conflicts with
the set of ongoing transactions in $\ongoingts[\tsvar][k]$
for each key $k$ written by $T'$
(\code{\ref{alg:aion}}{\ref{line:onlinechecksi-noconflictaxiom-conflict}}).
Note that we exclude $T'.\tid$ from $\ongoingts[\tsvar]$
at \code{\ref{alg:aion}}{\ref{line:onlinechecksi-ongoingts-t-setminus}}
to prevent reporting self-conflicts (i.e., $T'$ conflicting with itself).
This also prevents reporting duplicate conflicts:
in the case of conflicting transactions $T_{1}$ and $T_{2}$,
\onlinetool{} reports the conflict only once,
considering the transaction with the smaller commit timestamp.

\subsubsection*{Step \ding{174}~(\codes{\ref{alg:aion}}{\ref{line:onlinechecksi-step3-begin}}{\ref{line:onlinechecksi-step3-end}})}
To re-check the \extaxiom{} axiom for transactions
starting after $T$ commits,
\onlinetool{} iterates through timestamps $\tsvar$
greater than $T.\committime$ in ascending order
(\code{\ref{alg:aion}}{\ref{line:onlinechecksi-step3-begin}}).
If $\tsvar$ corresponds to the start timestamp of a transaction $T'$
\emph{and} the timeout for $T'$ has not expired,
\onlinetool{} re-checks the \extaxiom{} axiom
for each external read of $T'$ based on $\extval[T]$
(\code{\ref{alg:aion}}{\ref{line:onlinechecksi-extaxiom-ext-read}}),
instead of using the latest version of \frontierts{} before $\tsvar$
as in \emph{Step} \ding{172}.
This optimization is effective
because only the recently incoming transaction $T$
impacts the justification of external reads of $T'$.
For further optimization, \onlinetool{} re-checks
only the external reads of $T'$ on keys that are written by $T$
(\code{\ref{alg:aion}}{\ref{line:onlinechecksi-extaxiom-keys}} and
\code{\ref{alg:aion}}{\ref{line:onlinechecksi-extaxiom-ext-read}})
\emph{and} have not been overwritten by later transactions after $T$
(\code{\ref{alg:aion}}{\ref{line:onlinechecksi-extaxiom-keys-setminus}}).
Essentially, the values of these keys are still influenced by $T$.
Therefore, if $\tsvar$ corresponds to the commit timestamp of transaction $T'$,
it suffices for \onlinetool{} to update $\frontierts[\tsvar]$ only for these keys
(\code{\ref{alg:aion}}{\ref{line:onlinechecksi-extaxiom-frontierts}}).
The third optimization dictates that
once all keys written by $T$ are overwritten by later transactions,
the re-checking process for \extaxiom{} terminates
(\code{\ref{alg:aion}}{\ref{line:onlinechecksi-extaxiom-break}}).

The determination of violation or satisfaction of the \extaxiom{} axiom
at this stage is temporary and recorded in $T'.\extaxiom$
(\code{\ref{alg:aion}}{\ref{line:onlinechecksi-extaxiom-recheck-bot}}
and \code{\ref{alg:aion}}{\ref{line:onlinechecksi-extaxiom-recheck-top}}).
When the timeout for $T'$ expires,
\onlinetool{} reports an \extaxiom{} violation
if $T'.\extaxiom = \bot$
(\code{\ref{alg:aion}}{\ref{line:timeout-extaxiom-violation}}).

\subsubsection*{\gc~(\codes{\ref{alg:aion}}{\ref{line:online-proc-gc-begin}}{\ref{line:online-proc-gc-end}})}

However, in the worst-case scenario,
\onlinetool{} cannot safely recycle
any data structures or transactions due to asynchrony.
To mitigate memory usage,
\onlinetool{} performs garbage collection periodically and conservatively:
each time it transfers $\frontierts$, $\ongoingts$, and transactions
below a specified timestamp from memory to disk
(\codes{\ref{alg:aion}}{\ref{line:online-proc-gc-begin}}{\ref{line:online-proc-gc-end}},
marked as \savestyle).
\onlinetool{} reloads these data structures and transactions
as needed later on (refer to code lines marked as \loadstyle).
%%%%%%%%%%%%%%%%%%%%%%%%%

\subsubsection{Complexity and Correctness Analysis} \label{sss:online-complexity}

\onlinetool{} behaves similarly to \tool{} in terms of complexity
for (re-)checking the \sessionaxiom, \intaxiom, \extaxiom,
and \noconflictaxiom{} axioms when a new transaction is received.
However, instead of sorting transactions based on their timestamps \emph{a priori} as in \tool{},
\onlinetool{} inserts the new transaction into an already sorted list of transactions,
which can be done in logarithmic time using, for example, balanced binary search trees.
Therefore, the time complexity of \onlinetool{}
for each new transaction is $O(\log N + M)$,
where $N$ is the number of transactions received so far,
and $M$ is the number of operations in the transactions.

The space complexity of \onlinetool{} is primarily determined by
the memory required for storing the ever-growing history,
and the versioned $\frontierts$ and $\ongoingts$ data structures.
Thanks to its garbage collection mechanism,
when asynchrony is minimal,
\onlinetool{} only needs to keep in memory
a few recent transactions and versions of $\frontierts$ and $\ongoingts$.

% \revise{ \marginnote[R3.O1]{R3.O1}
\onlinetool{} follows a “simulate-and-check” approach, and its correctness is straightforward when all transactions are processed in ascending order based on their start/commit timestamps, ensuring no re-checking is required (by 
 the timeout mechanism described in Section \ref{ss:workflow}). However, when out-of-order transactions necessitate re-checking, it is crucial to determine which transactions should be re-checked against which axioms.
 % A formal argument is provided in \cite{aion_and_chronos}.
% }
%%%%%%%%%%%%%%%%%%%%%%%%%
% \subsubsection{Relations with \tool} \label{sss:online-offline}

% To mitigate the impact of asynchrony,
% transactions can be collected and processed in batches.
% Within each batch, \onlinetool{} can handle
% the start and commit events of transactions sequentially,
% following their ascending order.
% Even better, with bounded and predictable asynchrony,
% transactions can be batched in a way that
% eliminates the need for re-checking.
% In this scenario, \onlinetool{} behaves similarly to \tool{}.

% Alternatively, if all transactions are collected
% in ascending order based on their start or commit timestamps,
% the re-checking step for the \extaxiom{} axiom (Step \ding{174})
% in \onlinetool{} can be bypassed.
%%%%%%%%%%%%%%%%%%%%%%%%%
%%%%%%%%%%%%%%%%%%%%%%%%%%%%%%
%%%%%%%%%%%%%%%%%%%%%%%%%%%%%%

% We design our timestamp-based SI checking algorithms
% by first relating the operational semantics of SI and its axiomatic semantics,
% which tells what should an SI checking algorithm do.
% implementation.tex

\section{Implementation}
% timekiller-workflow.tex

\begin{figure}[t]
	\centering
	\includegraphics[width = 0.45\textwidth]{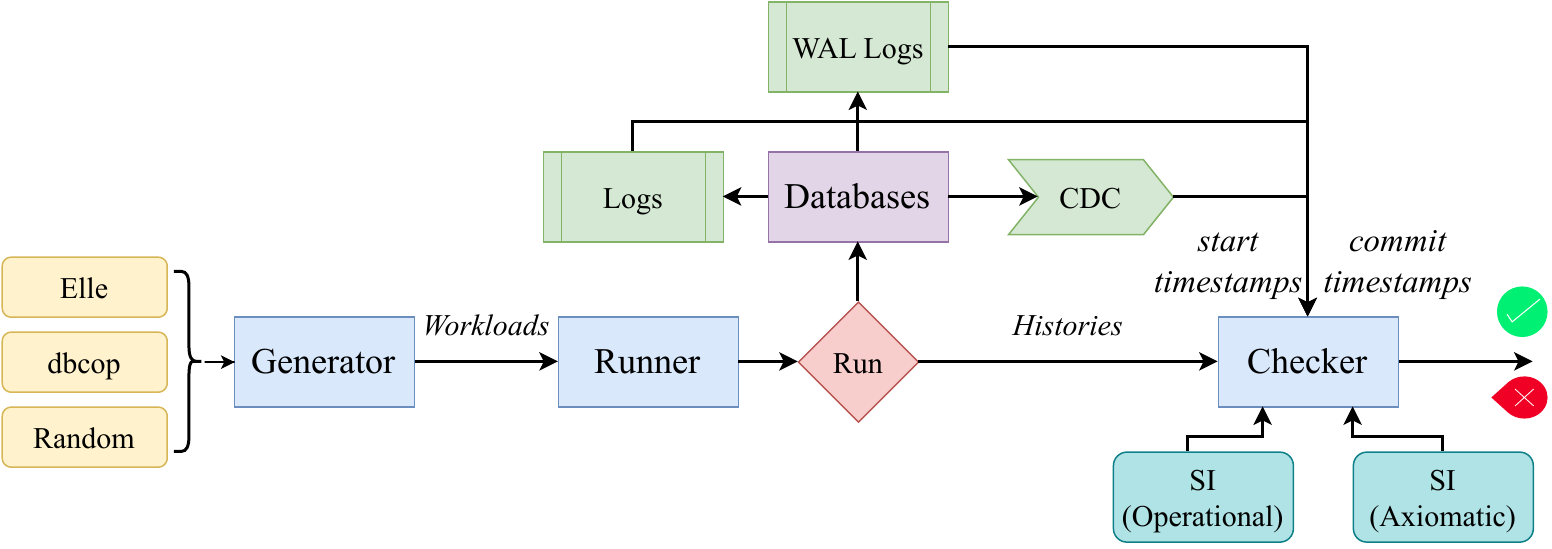}
	\caption{Workflow of online timestamp-based SI checking.}
	\label{fig:timekiller-workflow}
\end{figure}

We have implemented the \tool{} and \onlinetool{} checkers
in 3.6k lines of Java code, available at \cite{aion_and_chronos}.
% \footnote{Source code: \url{https://anonymous.4open.science/r/TimeKiller-243F}}
To evaluate performance,
we conducted extensive experiments
on three representative transactional databases:
the relational databases TiDB and YugabyteDB
% the document-oriented database MongoDB,
and the graph database Dgraph.
%%%%%%%%%%%%%%%
\subsection{Workflow} \label{ss:workflow}

Figure~\ref{fig:timekiller-workflow} illustrates
the workflow of online timestamp-based SI checking.
The generator creates a workload of transactions
based on specified templates and parameters,
which are executed by a database system to generate a history.
Our approach extracts the start and commit timestamps of each transaction 
from the database logs or its Change Data Capture (CDC) mechanism.
The history is then checked for violations.

% \revise{\marginnote[R3.O2]{R3.O2} 
The timeout mechanism sets a deadline for checking and re-checking the \extaxiom{} axioms for each transaction.
Within this period, \onlinetool{} does not report the \extaxiom{} violation as it is subject to change due to asynchrony.
Once the timeout expires, the \extaxiom{} checking result is considered final and is reported to clients.
In our evaluations, we conservatively set 5 seconds. 
% }
%%%%%%%%%%%%%%%
% \subsection{Deployment} \label{ss:deployment}

% \red{TODO: deployment; machine configuration}
%%%%%%%%%%%%%%%
\subsection{Generating Workloads}
\label{ss:generating-workloads}

To generate key-value histories,
we use various workloads tailored to the specific database systems:
\begin{itemize}
  \item For TiDB and YugabyteDB,
    we use a schema consisting of a two-column table,
    where one column stores keys and the other stores values.
    TiDB and YugabyteDB automatically translate the SQL statements into
    key-value operations on their underlying storage engines~\cite{TiDB:VLDB2020}.
  % \item For MongoDB, we use a simple document to represent a key-value pair.
  %     The implicit primary key field ``\_id'' serves as the key,
  %     while a new field ``v'' is added to represent the value.
  %     MongoDB automatically transforms operations on the document
  %     into key-value operations on its underlying storage engine, WiredTiger~\cite{TunableConsistency:VLDB2019}.
  \item For Dgraph, we represent key-value pairs as JSON-like graph nodes with two fields,
    ``uid'' and ``value''.
    Updates to node data are performed using ``mutation'' operations,
    which are transformed into key-value operations
    by Dgraph’s underlying storage engine.
\end{itemize}

  To generate list histories, we adapt our approach based on the data types.
  For example, in TiDB and YugabyteDB,
  a comma-separated TEXT field can be used to represent a list value.
  The append operation can be implemented using an
  ``\textsc{INSERT ... ON DUPLICATE KEY UPDATE}'' statement
  to either insert or concatenate values.
  % \revise{\marginnote{R3.D1}
  Following previous work~\cite{Elle:VLDB2020,Cobra:OSDI2020,PolySI:VLDB2023,Framework:CONCUR2015}, we consider only committed transactions for verification.
  % }
    % \item For MongoDB, we utilize its native array data type
    %   to represent lists.

% Note that although the checking algorithm \tool{}
% does not rely on the unique-value assumption,
% the history generator for some database system may need it.
% For example, \hfwei{TiDB}
% We ensure unique values written for each key using counters.
%%%%%%%%%%%%%%%
\subsection{Extracting Transaction Timestamps}
\label{ss:extracting-timestamps}

We summarize the lightweight methods
for extracting timestamps from the representative databases as follows:
(1)
% \revise{\marginpar[R1.D1]{R1.D1}
For \textbf{TiDB}, we modified its source code for CDC \cite{tidb_ts_get},
  % \footnote{TiCDC: https://github.com/pingcap/tiflow},
  adding simple print statements to expose transaction timestamps.
  % }
(2) \textbf{YugabyteDB} stores transaction timestamps 
    in the Write-Ahead Log (WAL).
  % \item MongoDB stores the start timestamps in log files
  %   and the commit timestamps in the special ``oplog.rs'' collection.
(3) For \textbf{Dgraph}, \emph{start} timestamps are included
    in the HTTP responses.
    We further modified Dgraph's source code to include
    \emph{commit} timestamps in HTTP responses as well,
    a change that has been officially merged into the main branch of Dgraph~\cite{pydgraph-pr}.

% \red{Note that} we set the commit timestamp of a read-only transaction
% to be the same as its start timestamp.
%%%%%%%%%%%%%%%%%%%%%%%%%%%%%%
\section{Experiments on \tool}
\label{section:offline-experiments}

In this section, we evaluate the offline checking algorithm \tool{}
and answer the following questions:

\begin{enumerate}[label=(\arabic*)]
  % \item Can the timestamp-based testing approach
  %   reveal SI violations in widely used production databases?
  \item How efficient is \tool, and how much memory does it consume?
    Can \tool{} outperform the state of the art under various workloads
    and scale up to large workloads?
  \item How do the individual components contribute to \tool's performance?
    Particularly, what impact does GC have
    on \tool's efficiency and memory usage?
  \item Can \tool{} effectively detect isolation violations?
\end{enumerate}
%%%%%%%%%%%%%%%%%%%%
% \input{sections/exp-si-violations}
% exp-perf.tex

%%%%%%%%%%%%%%%%%%%%%%%%%%%%%%
\subsection{Setup}\label{sss:perf-setup}

We conduct a comprehensive performance analysis of \tool,
comparing it to other checkers:
\textbf{PolySI}~\cite{PolySI:VLDB2023} and \textbf{Viper}~\cite{Viper:EuroSys2023} are both SI checkers designed for key-value histories. \textbf{Elle}~\cite{Elle:VLDB2020} serves as a checker for various isolation levels, including SER and SI. \textbf{Emme-SI}~\cite{Clark:EuroSys2024} is a version-order recovery-based SI checker.
\subsubsection{Workloads} \label{sss:perf-workloads}

% workload.tex

\begin{table}[t]
  \centering
  \caption{Parameters of default workload.}
  \label{table:workload-parameters}
  \renewcommand\arraystretch{1.2}
  \resizebox{\columnwidth}{!}{%
    \begin{tabular}{|c|c|c|}
      \hline
      \textbf{Parameters}   & \textbf{Values}  & \textbf{Default} \\ \hline
      \hline
      Number of sessions (\#sess)
          & $10$, $20$, $50$, $100$, $200$
          & $50$    \\ \hline
      Number of transactions (\#txns)
          & \incell{c}{$5$K, $100$K, $200$K, $500$K, $1, 000$K}
          & $100$K  \\ \hline
      \incell{c}{Number of operations per transaction\\(\#ops/txn)}
          & $5$, $15$, $30$, $50$, $100$
          & $15$    \\ \hline
      % \incell{c}{Ratio of read-only transactions (\%ro)}
      %     & $0\%$, $25\%$, $50\%$, $75\%$
      %     & $50\%$  \\ \hline
      \incell{c}{Ratio of read operations (\%reads)}
          & $10\%$, $30\%$, $50\%$, $70\%$, $90\%$
          & $50\%$  \\ \hline
      Number of keys (\#keys)                      & $200$, $500$, $1000$, $2000$, $5000$   & $1000$  \\ \hline
      Distribution of key access (dist)          & Uniform, Zipfian, Hotspot              & Zipfian \\ \hline
    \end{tabular}%
  }
\end{table}
(1) Default workloads.
We tune the seven workload parameters,
as shown in Table~\ref{table:workload-parameters},
during workload generation.
The ``Default'' column presents the default values for these parameters.
For the ``hotspot'' key-access distribution,
we mean that 80\% of operations target 20\% of keys.
% \revise{
(2)
% \marginpar[R1.O1]{R1.O1 R4.O2}
\textbf{Twitter} \cite{twitter_dataset} is a simple clone of Twitter. It allows users to create new tweets, follow/unfollow other accounts,
and view a timeline of recent tweets from those they follow.
In our study, we involved 500 users, each posting tweets of 140 words. %and follows or unfollows other users according to a Zipfian distribution ($\alpha$ = 100).
(3) \textbf{RUBiS} \cite{rubis_dataset} emulates an auction platform similar to eBay.
, allowing users to create accounts, list items, place bids, and leave comments.
We initialized the marketplace with 200 users and 800 items.
% }
%%%%%%%%%%%%%%%%%%%%
\subsubsection{Setup} 
% \revise{\marginnote[R4.O5]{R4.O5}
We evaluate all checkers using histories produced by industry databases. 
Specifically, for Twitter and RUBiS dataset, we collect SER and SI histories from Dgraph (v20.03.1). For default dataset,
we use SI histories collected from Dgraph (v20.03.1)
on a key-value store.
However, since Dgraph lacks support for list data types,
we use SI histories collected from TiDB (v7.1.0) 
to compare \tool{} with ElleList.
For SER checking, we collect histories from YugabyteDB (v2.20.7.1).
% }
Both Dgraph and TiDB are deployed on a cluster of 3 machines
in a local network.
Two of these machines are equipped with a 2.545GHz AMD EPYC 7K83 (2-core) processor
and 16GB memory,
while the third machine has a 2.25GHz AMD EPYC 9754 (16-core) processor and 64GB memory.
Both \tool{} and \onlinetool{} are deployed on the 16-core machine.
%%%%%%%%%%%%%%%%%%%%

\subsection{Performance and Scalability}
\label{ss:performance}
\subsubsection{Runtime and Scalability}
\label{sss:offline-time}

\begin{figure}[t]
	\centering
	\includegraphics[width = 0.25\textwidth]{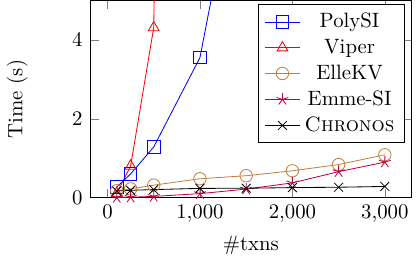}
 	\caption{Runtime comparison on key-value histories
 	  under varying number of transactions.}
 	\label{fig:exp-other-tools}
\end{figure}

% \begin{wrapfigure}{r}{0.22\textwidth}
% 	\includegraphics[width = 0.23\textwidth]{figs/exp-other-tools}
%	\caption{Runtime comparison on key-value histories
%	  under varying \#transactions.}
%	\label{fig:exp-other-tools}
% \end{wrapfigure}
% exp-other-tools-elle.tex

\begin{figure}[t]
	\begin{subfigure}{0.235\textwidth}
		\centering
		\includegraphics[width = \textwidth] {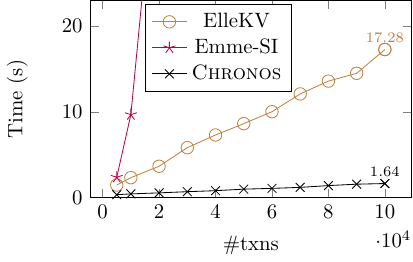}
		\caption{On key-value histories}
		\label{fig:exp-other-tools-ellekv}
	\end{subfigure}
	\begin{subfigure}{0.235\textwidth}
		\centering
		\includegraphics[width = \textwidth] {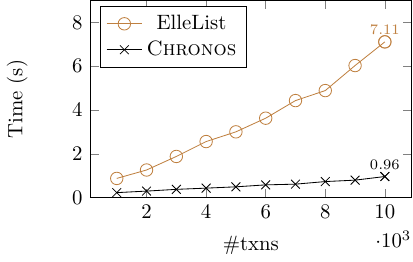}
		\caption{On list histories}
		\label{fig:exp-other-tools-ellelist}
	\end{subfigure}
	\caption{Runtime comparison with Emme-SI and ElleKV/ElleList
	  under varying number of transactions.}
	\label{fig:exp-other-tools-elle}
\end{figure}

As shown in Figure~\ref{fig:exp-other-tools},
\tool, ElleKV and Emme-SI, significantly outperform PolySI and Viper
on key-value histories,
which grow super-linearly with transaction numbers.
To further compare \tool{} with ElleKV and Emme-SI on larger key-value histories
and ElleList on list histories,
we increase the transactions
and plot the runtime in Figure~\ref{fig:exp-other-tools-ellekv}
and~\ref{fig:exp-other-tools-ellelist}, respectively.
Remarkably, \tool{} can check a key-value history with
100K transactions within 2 seconds
(Figure~\ref{fig:exp-other-tools-ellekv}).
Emme-SI performance is suboptimal due to its need to 
construct a start-ordered serialization graph for the entirety of the transaction history.
While ElleKV and ElleList show almost linear growth, \tool{} growth rate is much slower as it processes transactions in timestamp order without cycle detection.
Specifically, \tool{} is about 10.5x faster than ElleKV
and 7.4x faster than ElleList.
For list histories, which are more complex, \tool takes about one second to check a history with 10K transactions (Figure~\ref{fig:exp-other-tools-ellelist}).
% Emme-SI performs poorly as it requires constructing the start-ordered serialization graph 
%         for the entire history.
        
% exp-runtime.tex

\begin{figure}[t]
	\centering
	\includegraphics[width = 0.42\textwidth]{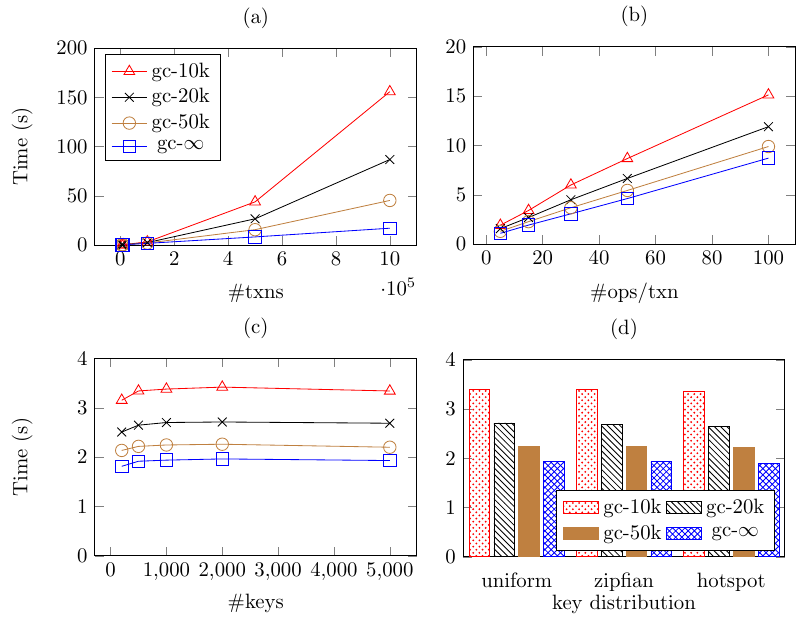}
	\caption{Runtime of \tool{} with various GC strategies
	  under varying workload parameters.}
	\label{fig:runtime}
\end{figure}

We further investigate the impact of varying workload parameters.
The results in Figure~\ref{fig:runtime}
{(marked with \blue{$\Box$})}
align with our time complexity analysis
in Section~\ref{sss:chronos-complexity}.
Specifically, the runtime of \tool{}
increases almost linearly with
both the total number of transactions (a)
and the number of operations per transaction (b),
while remaining stable for other parameters (c, d).
Notably in Figure~\ref{fig:runtime}(a),
\tool{} checks key-value histories with up to 1 million transactions in about 17 seconds, showing high scalability, whereas PolySI takes several hours and Viper struggles with large histories~\cite{PolySI:VLDB2023}.

% Figure~\ref{fig:runtime}a also illustrates that
% \tool{} can check key-value histories containing
% up to 1 million transactions (with default values for other parameters)
% in about 17 seconds,
% showcasing its remarkable scalability.
% Conversely, PolySI takes several hours % to check such large histories
% and Viper struggles to handle them effectively~\cite{PolySI:VLDB2023}.

In Figure~\ref{fig:runtime},
we also explore the impact of varying GC frequencies
on \tool's performance.
In these experiments, we trigger GC periodically after
processing a certain number of transactions
(e.g., 10K and 500K),
{with `gc-$\infty$' indicating that no GC is called at all}.
The results show that as GC is called more frequently,
\tool{} takes longer to check the history.
Moreover, while the runtime grows linearly with the number of operations per transaction (b), it exceeds linear growth concerning the total number of transactions (a) when GC is called more often, as GC frequencies are tuned based on transactions processed rather than operations. 
%Section~\ref{sss:decomposition} provides a deeper analysis of the time consumption for each stage of \tool{} under different GC frequencies.
%%%%%%%%%%%%%%%%%%%%
\subsubsection{Memory Usage and Scalability}
\label{sss:offline-memory}

% exp-memory.tex

\begin{figure}[t]
	\centering
	\includegraphics[width = 0.42\textwidth]{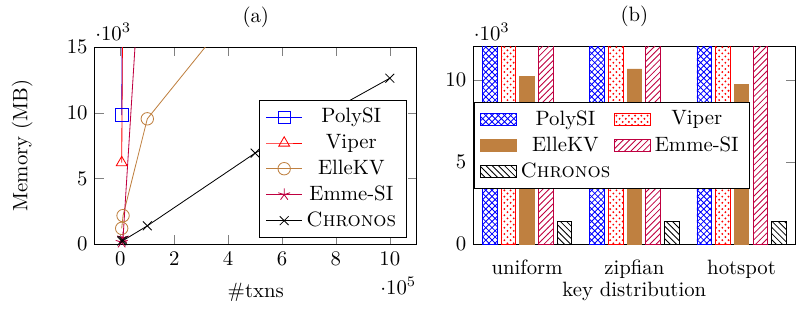}
	\caption{The maximum memory usage under varying workload parameters.}
	\label{fig:memory}
\end{figure}

As shown in Figure~\ref{fig:memory},
the \emph{maximum} memory usage of \tool{}
increases linearly with the number of transactions (a)
and the number of operations per transaction (b),
while it remains stable concerning other parameters
(omitted due to space limit).
This aligns with our space complexity analysis
in Section~\ref{sss:chronos-complexity}.
% In Section~\ref{sss:memory-vs-time},
% we further investigate the memory usage of \tool{} over time.

Figure~\ref{fig:memory}a shows that
\tool{} consumes about 13 GB of memory
for checking a key-value history with 1 million transactions.
In contrast, PolySI, Viper, and Emme-SI require
significantly more memory due to
additional polygraph/SSG structures~\cite{PolySI:VLDB2023}, while
ElleKV consumes more
due to its extensive dependency graphs~\cite{Elle:VLDB2020}.
%%%%%%%%%%%%%%%%%%%%
%%%%%%%%%%%%%%%%%%%%%%%%%%%%%%
% exp-closer.tex

%%%%%%%%%%%%%%%%%%%%%%%%%%%%%%
\subsection{A Closer Look at \tool}
\label{ss:closer}

%%%%%%%%%%%%%%%%%%%%
\subsubsection{Runtime Decomposition of \tool}
\label{sss:decomposition}

% exp-decompose.tex

\begin{figure}[t]
	\centering
	\includegraphics[width = 0.42\textwidth]{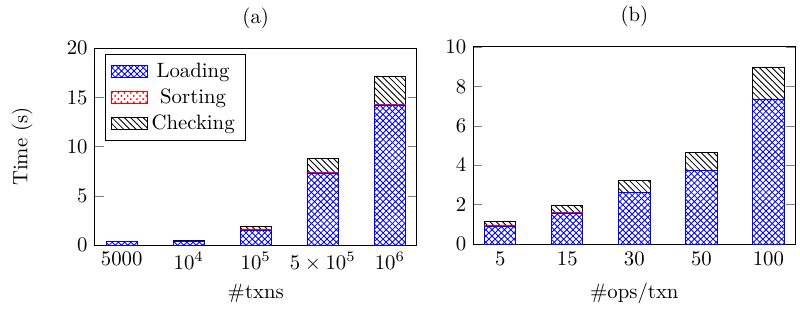}
	\caption{Runtime decomposition of \tool{} (without GC)
	  under varying workload parameters. }
	\label{fig:decompose}
\end{figure}

We measure \tool's checking time in terms of four stages:
\begin{itemize}
	\item \emph{loading} which loads the whole history into memory;
	\item \emph{sorting} which sorts the start and commit timestamps of transactions
	  in ascending order;
  \item \emph{checking} which checks the history by examing axioms;
  \item \emph{garbage collecting} (GC) which recycles transactions
    that are no longer needed during checking.
\end{itemize}

Figure~\ref{fig:decompose} shows the time consumption of each stage
of \tool{} \emph{without} GC under varying workload parameters.
First, the loading stage,
which involves frequent file I/O operations,
is the most time-consuming,
while the sorting stage is negligible.
Second, the time spent on both the loading and checking stages
increases almost linearly with
the total number of transactions (a) and the number of operations per transaction (b), remaining stable for other parameters (e.g., \#session, \#keys, read proportion, and key distribution).

Figure~\ref{fig:decompose-freq} shows that varying GC frequencies affect the time consumption of each stage
when checking a history of 1 million transactions.
Frequent GC calls make it the most time-consuming stage, while the time spent on GC decreases linearly as the frequency of GC decreases.
% The results indicate that GC becomes the most time-consuming stage
% when invoked frequently.
% The time spent on GC decreases linearly
% as the GC frequency decreases.
%%%%%%%%%%%%%%%%%%%%
\subsubsection{Memory Usage of \tool{} over Time}
\label{sss:memory-vs-time}

% exp-decompose-freq-and-memory-default.tex

\begin{figure}[t]
	\begin{minipage}{0.23\textwidth}
		\centering
		\includegraphics[width = 1\textwidth]{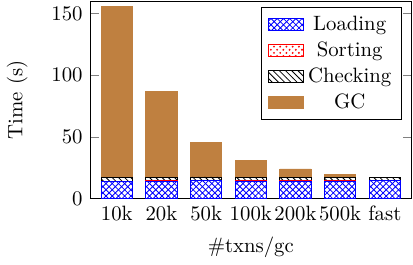}
		\caption{Runtime decomposition of \tool{} under varying GC frequencies.}
		\label{fig:decompose-freq}
	\end{minipage}
	\hfill
	\begin{minipage}{0.23\textwidth}
		\centering
		\includegraphics[width = 1\textwidth]{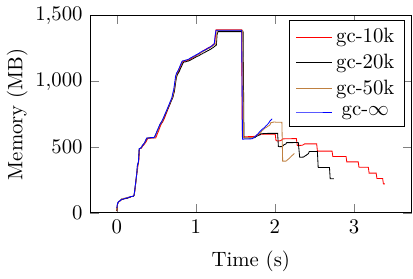}
		\caption{Memory usage of \tool{} over time.}
		\label{fig:memory-default}
	\end{minipage}
\end{figure}

Figure~\ref{fig:memory-default} depicts the memory usage of \tool{}
over time when checking a key-value history of 100K transactions.
Initially, memory usage increases steadily, peaking during the loading stage, followed by a sharp decline due to a JVM GC before checking. 
During the checking stage,
the memory usage displays a sawtooth pattern,
characterized by intermittent increases
followed by decreases after each GC.
Overall, the memory usage gradually decreases over time
until the checking stage ends.
% It is observed that
% more frequent GC calls result in smaller memory releases
% per GC and longer total runtime.
% Therefore, GC should be called judiciously in the offline checking stage.
More frequent GC calls lead to smaller memory releases per GC and longer total runtime, indicating that GC should be managed judiciously during the offline checking stage.
%%%%%%%%%%%%%%%%%%%%%%%%%%%%%%

\subsection{Checking Isolation Violations}
% \textcolor{red}{yugabyte dgraph bugs}
% \lhx{YugabyteDB 2.17.1.0 contains an issue related to clock skew, which may lead to violating \intaxiom{}. \tool{} is able to find the violations.}

\tool{} successfully reproduced a clock skew bug in YugabyteDB v2.17.1.0,
leading to\intaxiom{} violations under both SI and SER conditions.
Additionally, we injected timestamp-related faults
into the histories generated by Dgraph,
and \tool{} effectively detected these violations,
while non-timestamp-based tools failed.
This highlights an important advantage of the timestamp-based checking approach
over traditional black-box methods in terms of completeness~\cite{Clark:EuroSys2024}.

% ts-vs-nots.tex

\begin{figure}[t]
 	\centering
 	\includegraphics[width = 0.25\textwidth]{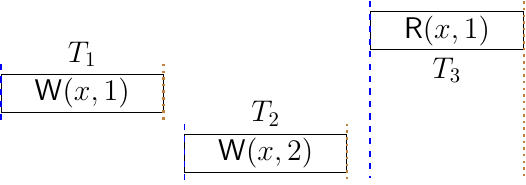}
 	\caption{A history that satisfies SI under traditional black-box SI checking
 	  but not under timestamp-based checking.}
 	\label{fig:ts-vs-nots}
\end{figure}

% \begin{wrapfigure}{r}{0.24\textwidth}
%	\includegraphics[width = 0.25\textwidth]{figs/ts-vs-nots}
%	\caption{A history that satisfies SI under traditional black-box SI checking
%	  but not under timestamp-based checking.}
%	\label{fig:ts-vs-nots}
% \end{wrapfigure}
For instance, in Figure~\ref{fig:ts-vs-nots},
where transactions $T_{1}$, $T_{2}$, and $T_{3}$
are committed sequentially.
Database developers might expect an SI violation
since $T_{3}$ reads from a snapshot that excludes the effect of $T_{2}$.
However, traditional non-timestamp-based SI checkers incorrectly infer an execution order of,
i.e., $T_{1}$, $T_{3}$, $T_{2}$, which did not occur.

% Note that this paper focuses not on discovering new bugs
% but on the design and implementation of efficient offline/online SER and SI checkers.
%%%%%%%%%%%%%%%%%%%%
%%%%%%%%%%%%%%%%%%%%%%%%%%%%%%
% experiments-online.tex

%%%%%%%%%%%%%%%%%%%%%%%%%%%%%%
\section{Experiments on \onlinetool}
\label{section:online-experiments}

In this section, we evaluate the online checking algorithm \onlinetool{}
and answer the following questions:

\begin{enumerate}[label=(\arabic*)]
  \item 
  What are the throughputs that \onlinetool{} (for SI checking) and \onlinetool-SER (for SER checking) can achieve?
  Can \onlinetool-SER outperform Cobra, the only existing online SER checker?
  \item How does the asynchronous arrival of transaction
    affect the stability of detecting the \extaxiom{} violations?
  \item How does the overhead of collecting histories
    affect the throughput of database systems? How does \onlinetool{} perform with constrained memory resources?
\end{enumerate}
%%%%%%%%%%%%%%%%%%%%
\subsection{Setup}
\label{ss:online-setup}

We generate transaction histories
using the workload described in Table~\ref{table:workload-parameters},
except for the throughput experiments (Section \ref{ss:throughput}),
where we use \#sess=24, \#ops/txn=8 for SER and SI checking, and \%reads=90\% for SER checking (to prevent Cobra from exceeding GPU memory).
The configuration of the database systems and the checkers is consistent with the setup described in Section~\ref{section:offline-experiments},
except for Cobra, which was tested using an NVIDIA GeForce RTX 3060 Ti GPU,
an AMD Ryzen 9 5900 (24-core) CPU, and 64GB of memory.
% For online checking, our setup comprises four machines,
% each equipped with a \red{2.2GHz Intel Xeon E5-2660 CPU (8 cores) and 64GB memory}.
% Three of these machines host the Dgraph (v20.03.1) cluster,
% which serves as the tested database,
% while the \onlinetool{} checker is deployed on a separate machine.
The network bandwidth between the node hosting \onlinetool{}
and the database instances is 20Mbps,
with an average latency of approximately 0.2ms.
History collectors dispatch transaction histories
to \onlinetool{} in batches of 500 transactions.
% \textcolor{red}{We continue to conduct an analysis of \onlinetool{}
% and compare it to an online SER checkers Cobra~\cite{Cobra:OSDI2020}.}
% \revise{
The \onlinetool-SER algorithm
% \marginnote[R4.O7]{R4.O7}
checks whether all transactions appear to execute sequentially in commit timestamp order.
% }
Note that start timestamps can be ignored,
and there is no need to check the \noconflictaxiom{} axiom.
% This makes SER checking more efficient than SI checking.
%, so start timestamps can be ignored and checking the \noconflictaxiom{} axiom is not required, making SER checking easier than SI.
% \revise{\marginnote[R4.O3, R4.O5]{R4.O3 R4.O5}
In our tests,
database throughput is lower than the checking speed
(see Sections \ref{ss:stability} and \ref{ss:online-collecting}).
To evaluate \onlinetool{}'s throughput limits,
we pre-collected logs and then fed historical data
exceeding the checkers' throughput
(see Section \ref{ss:throughput}).
% }
%%%%%%%%%%%%%%%%%%%%
\subsection{Throughput of Online Checking}
\label{ss:throughput}

% exp-online-tps-cobra.tex

\begin{figure}[t]
	\centering
	% \includegraphics[width = 0.30\textwidth]{figs/exp-online-tps-cobra.pdf}
	% \caption{Throughput of \onlinetool{} and Cobra over time.}
	% \label{fig:online-tps-cobra}
        \begin{subfigure}{0.235\textwidth}
		\centering
		\includegraphics[width = \textwidth] {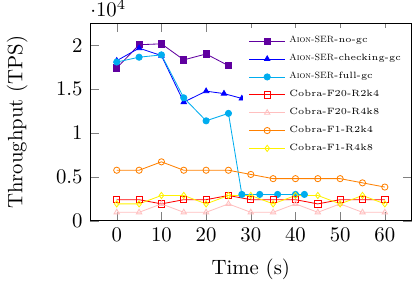}
		\caption{SER checking (Default)}
		\label{fig:online-tps-cobra}
	\end{subfigure}
        \begin{subfigure}{0.235\textwidth}
		\centering
		\includegraphics[width = \textwidth] {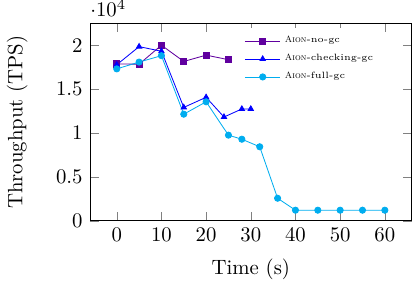}
		\caption{SI checking (Default)}
		\label{fig:online-tps-si-check}
	\end{subfigure}
        \begin{subfigure}{0.235\textwidth}
		\centering
		\includegraphics[width = \textwidth] {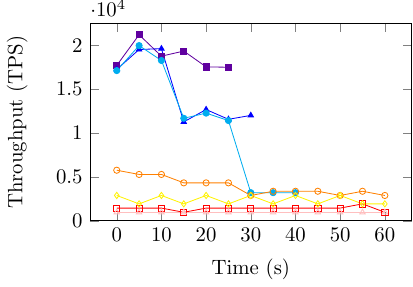}
		% \caption{\revise{SER checking (RUBiS)}}
            \caption{SER checking (RUBiS)}
		\label{fig:online-tps-rubis-ser}
	\end{subfigure}
 %        \begin{subfigure}{0.235\textwidth}
	% 	\centering
	% 	\includegraphics[width = \textwidth] {figs/exp-online-tps-rubis-si.pdf}
	% 	\caption{SI checking (RUBiS)}
	% 	\label{fig:online-tps-rubis-si}
	% \end{subfigure}
        \begin{subfigure}{0.235\textwidth}
		\centering
		\includegraphics[width = \textwidth] {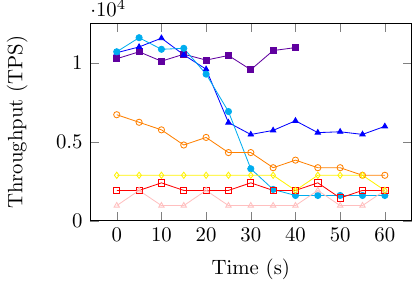}
		% \caption{\revise{SER checking (Twitter)}}
            \caption{SER checking (Twitter)}
		\label{fig:online-tps-twitter-ser}
	\end{subfigure}
 %        \begin{subfigure}{0.235\textwidth}
	% 	\centering
	% 	\includegraphics[width = \textwidth] {figs/exp-online-tps-twitter-si.pdf}
	% 	\caption{SI checking (Twitter)}
	% 	\label{fig:online-tps-twitter-si}
	% \end{subfigure}
    \caption{Throughput of online checking over time.}
    \label{fig:online-tps-throughput}
\end{figure}

% \textcolor{red}{replace to comparison to Cobra}
% Figure~\ref{fig:exp-online-tps} illustrates the throughput of \onlinetool{}
% alongside the database's throughput,
% measured in Transactions Per Second (TPS).
% In Figure~\ref{fig:exp-online-tps}(a), we set the number of operations
% per transaction to 8 to match the default setting of Cobra~\cite{Cobra:OSDI2020},
% an online checker for serializability
% (see Section~\ref{section:related-work} for comparison).
% The result demonstrates that \onlinetool{}
% can achieve a TPS of up to 4K.
% % sufficient to keep pace with the database's TPS.
% The throughput of \onlinetool{} has experienced some fluctuations
% due to periodic GC performed during the whole online checking stage.
% Nevertheless, the average TPS of \onlinetool{} in steady state
% is approximately 3K,
% sufficient to keep pace with the database's TPS.
% This indicates \onlinetool's capacity to effectively
% handle nearly 260 million transactions per day,
% approximately one third of the daily transaction volume
% of VISA's global payments and cash transactions in 2023~\cite{visa-fact-sheet}.  % VISA 750M txn per day
Figure~\ref{fig:online-tps-throughput} illustrates
the throughput of \onlinetool{} and \onlinetool-SER
under three different GC strategies,
as well as the throughput of Cobra
under varying fence frequencies (F) and round sizes (R).
The total number of transactions is 500K,
with an arrival rate exceeding 25K TPS for each checker.

For default benchmark, as shown in Figure \ref{fig:online-tps-cobra}, when no GC is performed, \onlinetool-SER achieves approximately 18K TPS. 
When GC is triggered
% \revise{
upon reaching a specified threshold 
for the total number of transactions in memory
% }
,
\onlinetool-SER's throughput decreases to about 12K TPS
and stabilizes at this level.
In scenarios where a maximum transaction limit is imposed on memory,
\onlinetool-SER executes GC immediately upon hitting this limit.
Since the transaction checking speed surpasses the GC process, 
\onlinetool-SER quickly reaches the limit again,
repeatedly triggering GC.
Under these conditions, the average throughput drops to roughly 3K TPS.

For Cobra, we tested both the default and optimal configurations
of fence frequency and round size,
as identified in~\cite{Cobra:OSDI2020}.
With a fixed fence frequency, a round size of 2.4K (the default value) transactions
consistently delivers superior performance.
In an extreme scenario where a fence is placed between every two transactions,
Cobra achieves a peak of around 6K TPS
during the first 25 seconds,
after which the throughput gradually declines.
In all other settings,
while Cobra's throughput remains stable over time,
it does not exceed 3K TPS.
% \input{figs/exp-online-tps-aionsi}

% \revise{\marginnote[R4.O6]{R4.O6}
We also used a history with 500K transactions generated under the SI level to test \onlinetool{}-SER.
The result shows that \onlinetool{} efficiently detects all 11839 violations with a checking speed comparable to that on violation-free histories, while Cobra terminates upon detecting the first violation.
We have validated the number of violations by \tool{}-SER.
% }

Figure~\ref{fig:online-tps-si-check} shows similar results for \onlinetool-SI.
Since more information must be stored for SI checking,
GC has a greater performance impact compared to SER checking.
Notably,
% \revise{
with no free memory
% }
,
\onlinetool{} maintains a throughput of approximately 1.2K TPS.
% \revise{ \marginpar[R1.O1]{R1.O1 R4.O2}
Figure~\ref{fig:online-tps-rubis-ser} and Figure~\ref{fig:online-tps-twitter-ser} show similar results across datasets.
Notably, \onlinetool's throughput decreases on the Twitter dataset compared to RUBiS,
due to the increasing number of keys,
requiring high computation and storage
to maintain the data structure $\frontierts$.
Cobra's performance depends on transaction dependencies,
influenced by key distribution, read-write ratios, and the number of operations per transaction,
rather than the number of keys.
% }

% \revise{\marginnote[R4.O4]{R4.O4}
While \onlinetool{} may not always keep up with peak TPS, it can exceed TPS during off-peak periods and eventually catch up. If the sustained TPS consistently exceeds \onlinetool{}’s checking speed, we recommend scaling resources or switching to an offline algorithm for efficient verification.
% }
% \blue{We extended the online checking process for a longer period (24 mins), during which the TPS of the database experienced two cycles of peak workloads.}
% In Figure~\ref{fig:exp-online-tps}(b),
% while maintaining the database's peak TPS at 4K,
% we increase the number of operations per transaction to 15.
% As expected, \onlinetool's TPS drops to approximately 2K
% due to the (linearly) increased runtime and space complexity
% of the checking algorithm.
% \onlinetool{} can continually process transactions at its maximum capacity.
% As the database's TPS decreases, \onlinetool{} gradually catches up.
%%%%%%%%%%%%%%%%%%%%
\subsection{Stability of Detecting the \extaxiom{} Violations}
\label{ss:stability}

We evaluate the impact of asynchrony
on the stability of detecting the \extaxiom{} violations
by counting the number of flip-flops,
denoting switches between $T.\extaxiom \gets \top$
and $T.\extaxiom \gets \bot$ for each transaction $T$.
As the history collector delivers transactions
to the checker in batches (500 transactions per batch),
we introduce artificial random delays
for each transaction within each batch,
following a normal distribution, to mimic asynchrony.
For these experiments,
we utilize workloads consisting of 10K transactions.

\begin{figure}[t]
	\centering
	\includegraphics[width = 0.45\textwidth]{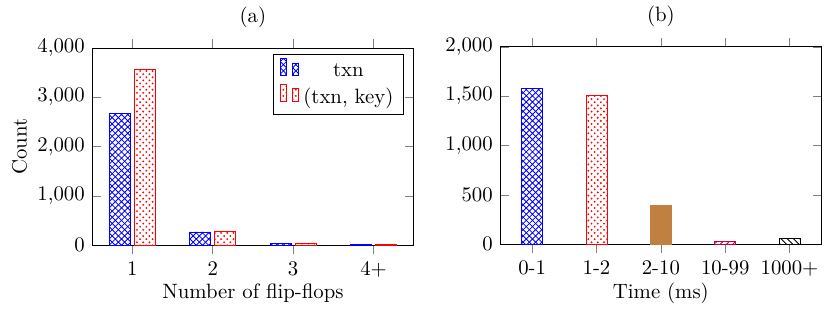}
	\caption{The number of flip-flops and
	  the time spent on rectifying the false positives/negatives.
		(The delays injected follow the normal distribution of $N(100,10^2)$.)}
	\label{fig:online-flip-100-10}
\end{figure}

An \extaxiom{} violation manifests as a transaction-key pair.
In Figure~\ref{fig:online-flip-100-10}(a),
the right bar denotes the number of \extaxiom{} violations,
while the left bar denotes the number of
unique transactions (txn) involved in these violations.
In these experiments, we inject delays
following a normal distribution with a mean of 100
and a standard deviation of 10, i.e., $N(100,10^2)$.
We observe that 29.8\% of transactions exhibit flip-flops,
with the vast majority (99\%) experiencing them once or twice.
Furthermore, Figure~\ref{fig:online-flip-100-10}(b) reveals that
over 95\% of the false positives/negatives are rectified within 10 ms.
Notably, approximately 3\% of the false positives/negatives
take about 1.5 seconds to be resolved,
probably in the subsequent batch.
% violations occurring within a time frame}
% of 100 milliseconds to 1 second
% are not resolved in the current batch,
% but are likely to be resolved during the subsequent batch (in \red{1.5s}).
Hence, setting a slightly higher time threshold
for reporting violations than the batch latency
would help mitigate the impact of these {false positives/negatives}.

\begin{figure}[t]
	\centering
	\includegraphics[width = 0.45\textwidth]{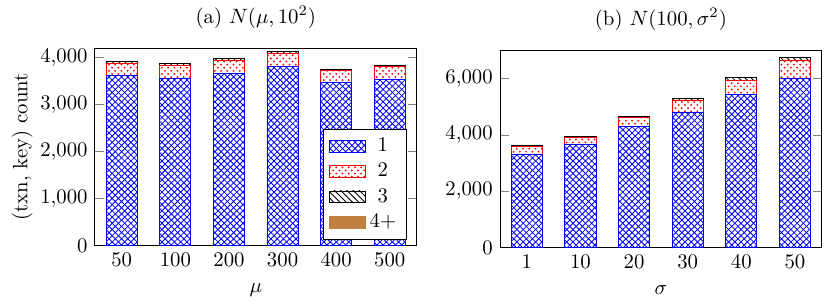}
  \caption{Number of flip-flops (in terms of (txn, key) count). }
	\label{fig:online-flip-compact}
    \vspace{-0.5cm}
\end{figure}

% We also explore the impact of the distribution of the delays
% on the number of flip-flops.
As shown in Figure~\ref{fig:online-flip-compact}(a),
the mean ($\mu$) of delays has a negligible impact
on the number of flip-flops
since transactions have been equally deferred.
Conversely, as shown in Figure~\ref{fig:online-flip-compact}(b),
increasing the standard deviation ($\sigma$) of delays
leads to a higher number of flip-flops,
% (\blue{we also observed a similar trend to the number of involved transactions in flip-flops})
due to the increased likelihood of transactions arriving out of order. %\blue{Similar to the impact on flip-flops, there are similar conclusions regarding the number of transactions involved in these flip-flops, which are greatly influenced by the standard deviation, as shown in Figure~\ref{fig:online-flip-total}.}

% \lhx{One may find that about 2,000 to 4,000 out of 10,000 transactions will give false violations, which is not a small percentage. But how long does it take to erase the violations? We investigate the 1-time flip-flops and the results are shown in Figure~\ref{fig:online-flip-time-mu} and Figure~\ref{fig:online-flip-time-sigma}.}

% \lhx{Of course, changing the mean of normal distribution does not make any difference at all. Most judgments are corrected within 1(2?) millisecond(s?). Over 95\% of the violations are eliminated within 10 milliseconds. When two out-of-order transactions are very close to each other, the latter can quickly correct the judgment errors caused when checking the former. An interesting finding is that no violations are removed between 100 milliseconds and 1 second. This is because a small part of violations still exist after all transactions in this batch are checked, and they are removed when checking the next batch. Considering the throughput, \onlinetool{} needs to wait over 1 second to receive the next request. The gap between 100 milliseconds and 1 second is when \onlinetool{} is waiting for the next request.}

% \lhx{Increasing the standard deviation makes \onlinetool{} need longer time to finalize the decisions on violations. Despite this, more than 95\% of violations are erased within 10 milliseconds.}

% \lhx{Through this experiment we can see that \onlinetool{} can correct most of the violations quickly, so although there may be a few false alarms it's not a big problem.}

%%%%%%%%%%%%%%%%%%%%
% \subsection{Overhead of Collecting Histories}
\subsection{Overhead}
\label{ss:online-collecting}

% \input{figs/exp-online-throughput-collection}
% \begin{figure}[t]
% 	\centering
% 	\includegraphics[width = 0.45\textwidth]{figs/exp-online-throughput-ops.pdf}
% 	\caption{Throughput with/without collecting history. \textcolor{red}{remove b c d}}
% 	\label{fig:online-throughput}
% \end{figure}

\begin{figure}[t]
	\begin{minipage}{0.23\textwidth}
		\centering
		\includegraphics[width = 1\textwidth]{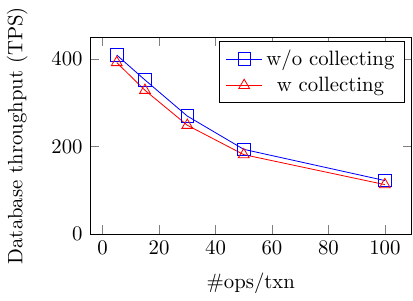}
	\caption{DB throughput with /without collecting history. }
	\label{fig:online-throughput}
	\end{minipage}
	\hfill
	\begin{minipage}{0.23\textwidth}
		\centering
		\includegraphics[width = 1\textwidth]{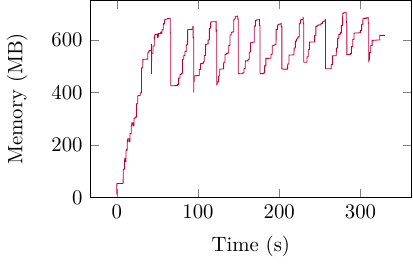}
	\caption{Constrained memory usage of \onlinetool{} over time.}
	\label{fig:exp-online-throughput-memory}
	\end{minipage}
    \vspace{-0.5cm}
\end{figure}

Figure~\ref{fig:online-throughput} compares
the database throughputs with and without history collection. 
%,
% under a workload comprising 20K transactions. % using default parameters.
% Across various parameters,
% (such as the number of operations per transaction,
% the number of sessions, the read/write ratio, and the number of keys),
We observe that collecting (and transmitting) the history
leads to a roughly 5\% decrease in TPS,
indicating a minor impact.
%%%%%%%%%%%%%%%%%%%%
%%%%%%%%%%%%%%%%%%%%%%%%%%%%%%
% The online GC of \onlinetool{} can be customized to
% enhance operational efficiency by setting minimum
% and maximum memory usage thresholds to trigger GC.
% It can also be triggered based on time intervals or completed TPS,
% although these aspects are not the primary focus of this paper.
% \input{figs/exp-online-throughput-memory}
% To evaluate the impact of memory constraints on \onlinetool{},
% we mandate that \onlinetool{} triggers GC
% when memory usage surpasses 700M
% in the experiment depicted
% in Figure~\ref{fig:exp-online-throughput-memory}.
% This experiment is conducted using a workload comprising 100K transactions.
% We observe that \onlinetool{} initially consumes memory gradually
% until it reaches the maximum memory usage threshold.
% Subsequently, memory usage oscillates between {400M and 700M}
% due to periodic GC,
% until the database finishes processing all transactions
% in approximately 310 seconds.
% Consequently, \onlinetool{} can effectively manage
% continuous online checking with relatively constrained memory resources.
In the experiment shown in Figure~\ref{fig:exp-online-throughput-memory}, we set \onlinetool{} to trigger garbage collection (GC) when memory usage exceeds 700M, using a workload of 100K transactions. Initially, memory consumption rises gradually until it hits the threshold, then oscillates between 400M and 700M due to periodic GC, completing all transactions in about 310 seconds. Thus, \onlinetool{} effectively manages continuous online checking with limited memory resources.
% \input{sections/discussions}
% related-work.tex

%%%%%%%%%%%%%%%%%%%%%%%%%%%%%%
\section{Related Work}
\label{section:related-work}

%%%%%%%%%%%%%%%%%%%%
% \paragraph*{SI in the Wild}

% Although SI is weaker than the gold standard of SER,
% it prevents a variety of common data anomalies from applications,
% including fractured reads, causality violation, lost update,
% and long fork~\cite{Framework:CONCUR2015}.
% According to~\cite{Responsibility:CIDR2023},
% 42 out of 93 ($45\%$) concurrency bugs
% % (in the ``Read followed by relevant Write'' category)
% collected from 46 different real-world open-source applications
% can be solved by upgrading the isolation level to SI.
% SI has been widely adopted by many commercial and open-source database systems.
% We have carefully examined \red{100} popular database systems
% selected from DB-Engines Ranking~\cite{DB-Engines},
% and found that \red{40} of them support SI.
% Readers can refer to our \red{online document~\cite{Isolation:VLDB2024}}
% for the detailed information of each database system.
% It is remarkable that \red{15} database systems opt for SI as a default isolation level,
% including \red{Memgraph, Dgraph}.
%%%%%%%%%%%%%%%%%%%%
The SER and SI checking problems are known to be \npc{}
for general key-value histories
with unknown transaction execution orders~\cite{SER:JACM1979, Complexity:OOPSLA2019}.
Existing checkers are offline
and fall into two categories:
those for general histories
and those for histories with additional information.

\paragraph*{Checkers for General Histories}

Algorithms for SER and SI checking, such as those in~\cite{Complexity:OOPSLA2019}, operate in polynomial time under fixed input parameters, but their efficiency is limited due to potentially large parameters~\cite{Cobra:OSDI2020, PolySI:VLDB2023}. Cobra encodes the SER problem into an SAT formula and solves it with the MonoSAT solver~\cite{MonoSAT:AAAI2015}. Similarly, PolySI encodes the SI problem into a SAT formula, also solved using MonoSAT. Viper~\cite{Viper:EuroSys2023} introduces BC-polygraphs to reduce the SI checking problem to cycle detection on BC-polygraphs, solved via MonoSAT. However, these checkers require exploring all potential transaction execution orders, leading to exponential computational costs with the number of transactions.

% The SER and SI checking algorithms proposed
% in~\cite{Complexity:OOPSLA2019} operate in polynomial time,
% assuming certain input history parameters,
% e.g., the number of sessions, are fixed.
% However, their practical efficiency is limited,
% as these fixed parameters can often be large~\cite{Cobra:OSDI2020, PolySI:VLDB2023}.
% Cobra~\cite{Cobra:OSDI2020} addresses the SER checking problem
% by encoding it into an SAT formula based on the polygraph-based
% characterization of SER
% and solving it with the MonoSAT solver~\cite{MonoSAT:AAAI2015}.
% Similarly, PolySI~\cite{PolySI:VLDB2023},
% influenced by both Cobra~\cite{Cobra:OSDI2020}
% and the dependency graph-based characterization of SI~\cite{AnalysingSI:JACM2018},
% encodes the SI checking problem into a SAT formula,
% also solved using MonoSAT.
% Viper~\cite{Viper:EuroSys2023} introduces BC-polygraphs
% as a novel representation of transactional dependencies,
% reducing the SI checking problem to
% a cycle detection problem on BC-polygraphs,
% which is then solved via MonoSAT.
% These checkers, however, require exploration of
% all potential transaction execution orders,
% resulting in computational costs that grow exponentially
% with the number $N$ of transactions in history. %in the worst case.
% % In contrast, our offline checking algorithm \tool{}
% % costs $O(N \log N + M)$ time,
% % where $M$ is the number of operations in the history.

\paragraph*{Checkers for Histories with Additional Information}

Elle \cite{Elle:VLDB2020} can infer transaction execution orders in histories with specific data types and the ``unique-value'' assumption, scaling well to tens of thousands of transactions but struggling with larger histories due to cycle detection costs. While Elle effectively handles certain data types, it has limited capabilities for others, such as key-value pairs. Clark et al.~\cite{SER-VO:ETH2021, Clark:EuroSys2024} introduced version order recovery, leading to the development of the timestamp-based checker Emme for SER and SI checking. However, Emme relies on costly cycle detection, making it unsuitable for online checking.

To our knowledge, no existing checkers support online SI checking like our \onlinetool{}. Cobra is the only checker that supports online SER checking but requires ``fence transactions'', which is often impractical in production environments.

\section{Conclusion, Discussions, and Future Work}
\label{section:conclusion}

In this work, we address the problem of online transactional isolation checking in database systems
by extending the timestamp-based offline isolation checking approach
to online settings.
Specifically, we develop an efficient offline SI checking algorithm,
\tool{}, which is inherently incremental.
We extend \tool{} to the online setting,
introducing the \onlinetool{} algorithm.
We also develop \onlinetool-SER, an online SER checker.
Experimental results demonstrate that
\onlinetool{} and \onlinetool-SER can handle online transactional workloads
with a sustained throughput of approximately 12K TPS.

% By leveraging the start and commit timestamps of transactions
% in a history, we can infer the transaction execution order
% and determine the snapshot of each transaction.
% which simulate the execution of transactions one by one
% and check the SI axioms on the fly.

% \revise{\marginnote[R4.O1, R4.07]{R4.O1 R4.O7}
Unlike black-box methods \cite{Cobra:OSDI2020,PolySI:VLDB2023},
\tool{} and \onlinetool{} require some
database knowledge and kernel modifications,
trading off for improved checking efficiency.
We plan to extend \tool{} and \onlinetool{}
to support more data types and complex queries,
such as SQL queries with predicates,
by inferring the commit/version order \cite{Clark:EuroSys2024}
and/or transaction snapshots from their start and commit timestamps/IDs.
% }

% \revise{
% \marginnote{R3.O3}
Our work do not directly work for database systems 
which implement SI/SER using mechanisms distinct from the timestamp-based methods.
However, our approach can be generalized to handle more database systems.
For example, Emme \cite{Clark:EuroSys2024} leverages PostgreSQL’s logical streaming replication and the Debezium's CDC tool to recover the version order, which is defined as the commit order of transactions. %We will explore these two methods in the future work.
In their SI implemenations, databases like SQL Server \cite{SQLServer}, PostgreSQL \cite{PostgreSQL}, and WiredTiger \cite{WiredTiger-Transaction} rely on the visibility rules to compute a transaction’s snapshot. They associate each transaction with a unique ID, but the transaction IDs, such as those used in WiredTiger, may not directly correlate with the timestamps employed in the database systems considered in this paper. We will explore how to obtain the snapshots in a lightweight way in these databases in future work.
% } 

% \revise{\marginnote[R3.D2]{R3.D2} 
The intra-transaction savepoints~\cite{DBLP:books/mk/WeikumV2002:book} are useful for partial rollbacks which can be supported by the underlying recovery algorithm.
However, we do not test the recovery mechanism
and simply assume it works correctly in this paper
and plan to address this in future work.
% }
% \red{few sentences for limitation of queries and databases that are not applicable}
%and graph path queries.
% To further enhance \onlinetool's scalability,
% we aim to explore parallelization techniques
% and the use of GPUs to accelerate the checking process.

% i.e., the set of transaction identifiers visible to it.

% These systems assign each transaction a unique identifier
% that does not directly correspond to timestamps.
% For instance, transactions are not necessarily
% committed in the order of their identifiers.
% Moreover, conditions like $T_{1}.\tid < T_{2}.\tid$
% and $T_{2}$ being visible to $T_{3}$
% do not necessarily imply that $T_{1}$ is also visible to $T_{3}$.

\section*{Acknowledgment}

We thank the anonymous reviewers for their valuable feedback.
This work is supported by NSFC (62472214),
Natural Science Foundation of Jiangsu Province (BK20242014).
Hengfeng Wei is the corresponding author.

\balance

\bibliographystyle{IEEEtran}
%\bibliography{si-checker}

\begin{thebibliography}{10}
\providecommand{\url}[1]{#1}
\csname url@samestyle\endcsname
\providecommand{\newblock}{\relax}
\providecommand{\bibinfo}[2]{#2}
\providecommand{\BIBentrySTDinterwordspacing}{\spaceskip=0pt\relax}
\providecommand{\BIBentryALTinterwordstretchfactor}{4}
\providecommand{\BIBentryALTinterwordspacing}{\spaceskip=\fontdimen2\font plus
\BIBentryALTinterwordstretchfactor\fontdimen3\font minus \fontdimen4\font\relax}
\providecommand{\BIBforeignlanguage}[2]{{%
\expandafter\ifx\csname l@#1\endcsname\relax
\typeout{** WARNING: IEEEtran.bst: No hyphenation pattern has been}%
\typeout{** loaded for the language `#1'. Using the pattern for}%
\typeout{** the default language instead.}%
\else
\language=\csname l@#1\endcsname
\fi
#2}}
\providecommand{\BIBdecl}{\relax}
\BIBdecl

\bibitem{SER:JACM1979}
\BIBentryALTinterwordspacing
C.~H. Papadimitriou, ``The serializability of concurrent database updates,'' \emph{J. ACM}, vol.~26, no.~4, pp. 631--653, oct 1979. [Online]. Available: \url{https://doi.org/10.1145/322154.322158}
\BIBentrySTDinterwordspacing

\bibitem{Bernstein:Book1986}
P.~A. Bernstein, V.~Hadzilacos, and N.~Goodman, \emph{{Concurrency Control and Recovery in Database Systems}}.\hskip 1em plus 0.5em minus 0.4em\relax USA: Addison-Wesley Longman Publishing Co., Inc., 1986.

\bibitem{PostgreSQL}
{PostgreSQL}, ``Database concurrency in {PostgreSQL},'' \url{https://www.red-gate.com/simple-talk/databases/postgresql/database-concurrency-in-postgresql/}.

\bibitem{CockroachDB}
{CockroachDB}, \url{https://www.cockroachlabs.com/product/}.

\bibitem{YugabyteDB}
YugabyteDB, \url{https://www.yugabyte.com/}.

\bibitem{HAT:VLDB2013}
\BIBentryALTinterwordspacing
P.~Bailis, A.~Davidson, A.~Fekete, A.~Ghodsi, J.~M. Hellerstein, and I.~Stoica, ``Highly available transactions: Virtues and limitations,'' \emph{Proc. VLDB Endow.}, vol.~7, no.~3, pp. 181--192, nov 2013. [Online]. Available: \url{https://doi.org/10.14778/2732232.2732237}
\BIBentrySTDinterwordspacing

\bibitem{CritiqueANSI:SIGMOD1995}
\BIBentryALTinterwordspacing
H.~Berenson, P.~Bernstein, J.~Gray, J.~Melton, E.~O'Neil, and P.~O'Neil, ``A critique of {ANSI SQL} isolation levels,'' in \emph{SIGMOD '95}.\hskip 1em plus 0.5em minus 0.4em\relax ACM, 1995, pp. 1--10. [Online]. Available: \url{https://doi.org/10.1145/223784.223785}
\BIBentrySTDinterwordspacing

\bibitem{dgraph}
Dgraph, \url{https://dgraph.io/}.

\bibitem{WiredTiger-Transaction}
``{WiredTiger} (version 10.0.2): Snapshot,'' \url{http://source.wiredtiger.com/mongodb-5.0/arch-snapshot.html}.

\bibitem{Percolator:OSDI2010}
D.~Peng and F.~Dabek, ``Large-scale incremental processing using distributed transactions and notifications,'' in \emph{OSDI'10}.\hskip 1em plus 0.5em minus 0.4em\relax USA: USENIX Association, 2010, pp. 251--264.

\bibitem{MongoDB}
MongoDB, \url{https://www.mongodb.com/}.

\bibitem{TiDB}
{TiDB}, \url{https://en.pingcap.com/tidb/}.

\bibitem{PostgreSQL-Jepsen}
{Jepsen testing of PostgreSQL 12.3}, \url{https://jepsen.io/analyses/postgresql-12.3}.

\bibitem{MongoDB-Jepsen}
{Jepsen testing of {MongoDB} 4.2.6}, \url{http://jepsen.io/analyses/mongodb-4.2.6}.

\bibitem{TiDB-Jepsen}
{Jepsen testing of {TiDB} 2.1.7}, \url{https://jepsen.io/analyses/tidb-2.1.7}.

\bibitem{Cobra:OSDI2020}
C.~Tan, C.~Zhao, S.~Mu, and M.~Walfish, ``{Cobra}: Making transactional key-value stores verifiably serializable,'' in \emph{OSDI'20}, 2020.

\bibitem{Complexity:OOPSLA2019}
\BIBentryALTinterwordspacing
R.~Biswas and C.~Enea, ``On the complexity of checking transactional consistency,'' \emph{Proc. ACM Program. Lang.}, vol.~3, no. OOPSLA, Oct. 2019. [Online]. Available: \url{https://doi.org/10.1145/3360591}
\BIBentrySTDinterwordspacing

\bibitem{CockroachDB-Jepsen}
{Jepsen testing of {CockroachDB} beta-20160829}, \url{https://jepsen.io/analyses/cockroachdb-beta-20160829}.

\bibitem{Elle:VLDB2020}
K.~Kingsbury and P.~Alvaro, ``{Elle}: Inferring isolation anomalies from experimental observations,'' \emph{Proc. VLDB Endow.}, vol.~14, no.~3, pp. 268--280, Nov. 2020.

\bibitem{PolySI:VLDB2023}
\BIBentryALTinterwordspacing
K.~Huang, S.~Liu, Z.~Chen, H.~Wei, D.~Basin, H.~Li, and A.~Pan, ``Efficient black-box checking of snapshot isolation in databases,'' \emph{Proc. VLDB Endow.}, vol.~16, no.~6, p. 1264–1276, apr 2023. [Online]. Available: \url{https://doi.org/10.14778/3583140.3583145}
\BIBentrySTDinterwordspacing

\bibitem{Viper:EuroSys2023}
\BIBentryALTinterwordspacing
J.~Zhang, Y.~Ji, S.~Mu, and C.~Tan, ``Viper: A fast snapshot isolation checker,'' in \emph{Proceedings of the Eighteenth European Conference on Computer Systems}, ser. EuroSys '23.\hskip 1em plus 0.5em minus 0.4em\relax New York, NY, USA: Association for Computing Machinery, 2023, p. 654–671. [Online]. Available: \url{https://doi.org/10.1145/3552326.3567492}
\BIBentrySTDinterwordspacing

\bibitem{DBLP:journals/pvldb/GuLXWCB24}
L.~Gu, S.~Liu, T.~Xing, H.~Wei, Y.~Chen, and D.~A. Basin, ``Isovista: Black-box checking database isolation guarantees,'' \emph{Proc. {VLDB} Endow.}, vol.~17, no.~12, pp. 4325--4328, 2024.

\bibitem{VerifyingMongoDB:arXiv2022}
H.~Ouyang, H.~Wei, Y.~Huang, H.~Li, and A.~Pan, ``Verifying transactional consistency of mongodb,'' 2022.

\bibitem{SER-VO:ETH2021}
J.~Clark, ``Verifying serializability protocols with version order recovery,'' Master's thesis, ETH Zurich, 2021, \url{https://doi.org/10.3929/ethz-b-000507577}.

\bibitem{Clark:EuroSys2024}
\BIBentryALTinterwordspacing
J.~Clark, A.~F. Donaldson, J.~Wickerson, and M.~Rigger, ``Validating database system isolation level implementations with version certificate recovery,'' in \emph{Proceedings of the Nineteenth European Conference on Computer Systems}, ser. EuroSys '24, 2024, p. 754–768. [Online]. Available: \url{https://doi.org/10.1145/3627703.3650080}
\BIBentrySTDinterwordspacing

\bibitem{Adya:PhDThesis1999}
A.~Adya, ``Weak consistency: A generalized theory and optimistic implementations for distributed transactions,'' Ph.D. dissertation, USA, 1999.

\bibitem{PSI:SOSP2011}
\BIBentryALTinterwordspacing
Y.~Sovran, R.~Power, M.~K. Aguilera, and J.~Li, ``Transactional storage for geo-replicated systems,'' in \emph{SOSP '11}.\hskip 1em plus 0.5em minus 0.4em\relax ACM, 2011, pp. 385--400. [Online]. Available: \url{https://doi.org/10.1145/2043556.2043592}
\BIBentrySTDinterwordspacing

\bibitem{Framework:CONCUR2015}
A.~Cerone, G.~Bernardi, and A.~Gotsman, ``A framework for transactional consistency models with atomic visibility,'' in \emph{CONCUR'15}, ser. LIPIcs, vol.~42.\hskip 1em plus 0.5em minus 0.4em\relax Schloss Dagstuhl--Leibniz-Zentrum fuer Informatik, 2015, pp. 58--71.

\bibitem{AnalysingSI:JACM2018}
\BIBentryALTinterwordspacing
A.~Cerone and A.~Gotsman, ``Analysing snapshot isolation,'' \emph{J. ACM}, vol.~65, no.~2, Jan. 2018. [Online]. Available: \url{https://doi.org/10.1145/3152396}
\BIBentrySTDinterwordspacing

\bibitem{MonkeyDB:OOPSLA2021}
\BIBentryALTinterwordspacing
R.~Biswas, D.~Kakwani, J.~Vedurada, C.~Enea, and A.~Lal, ``{MonkeyDB}: Effectively testing correctness under weak isolation levels,'' \emph{Proc. ACM Program. Lang.}, vol.~5, no. OOPSLA, oct 2021. [Online]. Available: \url{https://doi.org/10.1145/3485546}
\BIBentrySTDinterwordspacing

\bibitem{LazyReplSI:VLDB2006}
K.~Daudjee and K.~Salem, ``Lazy database replication with snapshot isolation,'' in \emph{VLDB'06}.\hskip 1em plus 0.5em minus 0.4em\relax VLDB Endowment, 2006, pp. 715--726.

\bibitem{aion_and_chronos}
\emph{\textsc{Aion} and \textsc{Chronos}}, March 2025, \url{https://github.com/FertileFragrance/TimeKiller}.

\bibitem{TiDB:VLDB2020}
\BIBentryALTinterwordspacing
D.~Huang, Q.~Liu, Q.~Cui, Z.~Fang, X.~Ma, F.~Xu, L.~Shen, L.~Tang, Y.~Zhou, M.~Huang, W.~Wei, C.~Liu, J.~Zhang, J.~Li, X.~Wu, L.~Song, R.~Sun, S.~Yu, L.~Zhao, N.~Cameron, L.~Pei, and X.~Tang, ``{TiDB}: A raft-based {HTAP} database,'' \emph{Proc. VLDB Endow.}, vol.~13, no.~12, p. 3072–3084, aug 2020. [Online]. Available: \url{https://doi.org/10.14778/3415478.3415535}
\BIBentrySTDinterwordspacing

\bibitem{tidb_ts_get}
Tsunaou, \emph{TiDB timestamp acquisition}, February 2025, \url{https://github.com/Tsunaou/tiflow/commit/82446cd9bd8a2feb52a7a3f403039620aaec9618}.

\bibitem{pydgraph-pr}
{pydgraph issue}, ``{return `commit\_ts' in the function commit() in txn.py (\#213)},'' \url{https://github.com/dgraph-io/pydgraph/pull/213}.

\bibitem{twitter_dataset}
Twitter, \emph{Big Data in Real Time at Twitter}, February 2025, \url{https://www.infoq.com/presentations/Big-Data-in-Real-Time-at-Twitter/ }.

\bibitem{rubis_dataset}
sguazt, \emph{Rice University Bidding System (RUBiS)}, February 2025, \url{https://github.com/sguazt/RUBiS}.

\bibitem{MonoSAT:AAAI2015}
S.~Bayless, N.~Bayless, H.~H. Hoos, and A.~J. Hu, ``{SAT} modulo monotonic theories,'' in \emph{Proceedings of the Twenty-Ninth AAAI Conference on Artificial Intelligence}, ser. AAAI'15, 2015, pp. 3702--3709.

\bibitem{SQLServer}
{Microsoft SQL Server}, ``Snapshot isolation in {SQL Server},'' Accessed April, 2024, \url{https://learn.microsoft.com/en-us/dotnet/framework/data/adonet/sql/snapshot-isolation-in-sql-server}.

\bibitem{DBLP:books/mk/WeikumV2002:book}
G.~Weikum and G.~Vossen, \emph{Transactional Information Systems: Theory, Algorithms, and the Practice of Concurrency Control and Recovery}.\hskip 1em plus 0.5em minus 0.4em\relax Morgan Kaufmann, 2002.

\bibitem{Raft:ATC2014}
D.~Ongaro and J.~Ousterhout, ``In search of an understandable consensus algorithm,'' in \emph{Proceedings of the 2014 USENIX Conference on USENIX Annual Technical Conference}, ser. USENIX ATC'14.\hskip 1em plus 0.5em minus 0.4em\relax USA: USENIX Association, 2014, p. 305–320.

\bibitem{ClockSI:SRDS2013}
\BIBentryALTinterwordspacing
J.~Du, S.~Elnikety, and W.~Zwaenepoel, ``{Clock-SI}: Snapshot isolation for partitioned data stores using loosely synchronized clocks,'' in \emph{SRDS '13}.\hskip 1em plus 0.5em minus 0.4em\relax USA: IEEE Computer Society, 2013, pp. 173--184. [Online]. Available: \url{https://doi.org/10.1109/SRDS.2013.26}
\BIBentrySTDinterwordspacing

\bibitem{Dgraph:TR2020}
\BIBentryALTinterwordspacing
M.~Jain, ``Dgraph: Synchronously replicated, transactional and distributed graph database,'' Dgraph Labs, Inc., Technical Report, February 2020. [Online]. Available: \url{https://dogy.io/wp-content/uploads/2021/04/dgraph.pdf}
\BIBentrySTDinterwordspacing

\bibitem{Spanner:OSDI2012}
J.~C. Corbett, J.~Dean, M.~Epstein, A.~Fikes, C.~Frost, J.~J. Furman, S.~Ghemawat, A.~Gubarev, C.~Heiser, P.~Hochschild, W.~Hsieh, S.~Kanthak, E.~Kogan, H.~Li, A.~Lloyd, S.~Melnik, D.~Mwaura, D.~Nagle, S.~Quinlan, R.~Rao, L.~Rolig, Y.~Saito, M.~Szymaniak, C.~Taylor, R.~Wang, and D.~Woodford, ``Spanner: {Google}'s globally-distributed database,'' in \emph{Proceedings of the 10th USENIX Conference on Operating Systems Design and Implementation}, ser. OSDI'12.\hskip 1em plus 0.5em minus 0.4em\relax USA: USENIX Association, 2012, p. 251–264.

\bibitem{Spanner:TOCS2013}
\BIBentryALTinterwordspacing
------, ``Spanner: {Google}'s globally distributed database,'' \emph{ACM Trans. Comput. Syst.}, vol.~31, no.~3, aug 2013. [Online]. Available: \url{https://doi.org/10.1145/2491245}
\BIBentrySTDinterwordspacing

\end{thebibliography}

%%%% appendix
\clearpage
\appendix
% ts-impl.tex
\nobalance
%%%%%%%%%%%%%%%%%%%%%%%%%%%%%%
% \subsection{Empirical Study on timestamp-based SI Implementations}
% \label{section:ts-impl}

% \input{tables/si-impl}

In this section, we delve into the timestamp-based SI implementations
of three representative transactional database systems:
the relational databases TiDB and YugabyteDB,
% the document-oriented database MongoDB,
and the graph database Dgraph.
% This justifies the feasibility of
% the timestamp-based SI checking approach. 
We also provide more information for extra experiments and the correctness of \onlinetool{}. 

%%%%%%%%%%%%%%%%%%%%%%%%%
\subsection{Centralized vs. Decentralized Timestamping}
\label{ss:timestamping}

Timestamp-based SI implementations are challenging in distributed settings,
particularly when a distributed transaction spans multiple nodes.
Two prevalent timestamping mechanisms in distributed settings
are centralized timestamping and decentralized timestamping.

In the centralized timestamping mechanism,
a centralized timestamp oracle is responsible for
providing strictly increasing timestamps for transactions.
To enhance robustness, this centralized timestamp oracle
is often replicated, such as in a Raft~\cite{Raft:ATC2014} group.

On the other hand, in the decentralized timestamping mechanism,
timestamps are generated by different nodes that are loosely synchronized.
These timestamps are not guaranteed to be
issued in a strictly increasing order.
This gives rise to the issue of ``snapshot unavailability''~\cite{ClockSI:SRDS2013},
where the snapshot specified by the start timestamp of a transaction
might not be readily available on certain participant nodes of the transaction.
%%%%%%%%%%%%%%%%%%%%%%%%%
\subsection{Timestamp-based SI Implementations}
\label{ss:ts-impl}

Databases that employ \emph{centralized timestamping} mechanism
include TiDB and Dgraph.
% and Apache Omid.
%%%%%%%%%%%%%%%%%%%%
\subsubsection{TiDB} \label{sss:tidb}

In TiDB, alongside a distributed storage layer (e.g., TiKV)
and a computation SQL engine layer,
there is a Placement Driver (PD) serving as a timestamp oracle.
The utilization of start and commit timestamps
during transaction processing in TiDB
is briefly described as follows~\cite{TiDB:VLDB2020}:

\begin{itemize}
	\item To begin a transaction,
	  the SQL engine asks PD for the start timestamp of the transaction.
	\item The SQL engine executes SQL statements by reading data from
		TiKV and writing to a local buffer.
		The snapshot provided by TiKV is taken at the most recent commit timestamp
		before the transaction's start timestamp.
	\item To commit a transaction, the SQL engine starts the 2PC protocol
	  and asks PD for the commit timestamp of the transaction when necessary,
		following the approach of Google Percolator~\cite{Percolator:OSDI2010}.
\end{itemize}
	% \item To commit a transaction, the SQL engine starts the 2PC protocol.
	%   It first performs the prewrite phase on the chosen primary key
	% 	with TiKV nodes.
	% \item If all prewrites succeed, the SQL engine asks PD for the commit timestamp of the transaction,
	% 	and performs the commit phase on the primary key with TiKV nodes.
	%   Note that the SQL engine commits the secondary keys asynchronously
	% 	with TiKV nodes.
%%%%%%%%%%%%%%%%%%%%
\subsubsection{Dgraph} \label{ss:dgraph}

Dgraph, a distributed GraphQL database,
supports full ACID-compliant cluster-wide distributed transactions.
The Zero group of a Dgraph cluster runs an oracle
which hands out monotonically increasing logical timestamps
for transactions in the cluster~\cite{Dgraph:TR2020}.
It is guaranteed that~\cite{Dgraph:TR2020},
for any transactions $T_{i}$ and $T_{j}$,
	(a) if $T_{i}$ commits before $T_{j}$ starts,
	  then $T_{i}.\committime < T_{j}.\starttime$;
	(b) if $T_{i}.\committime < T_{j}.\starttime$,
	  then the effects of $T_{i}$ are visible to $T_{j}$; \emph{and}
	(c) if $T_{i}$ commits before $T_{j}$ commits,
	  then $T_{i}.\committime < T_{j}.\committime$.
%%%%%%%%%%%%%%%%%%%%

Databases that employ \emph{decentralized timestamping} mechanism
include MongoDB and YugabyteDB.

% \subsubsection{MongoDB} \label{sss:mongodb}

% MongoDB deployment is a sharded cluster, a replica set,
% or standalone~\cite{TunableConsistency:VLDB2019}.
% A standalone is a storage node that represents a single instance of a data store.
% A replica set consists of a primary node and several secondary nodes.
% A sharded cluster is a group of multiple replica sets, among which data is sharded.
% We focus on the SI implementations in a replica set and in a sharded cluster.

% In a replica set, the primary node maintains an oplog of transactions,
% where each entry is assigned a unique commit timestamp (i.e., HLC).
% These commit timestamps determine the (logical) commit order of transactions.
% When a transaction starts, it is assigned a {read timestamp} on the primary
% such that all transactions with smaller commit timestamps have been (physically) committed.
% % in the underlying storage engine.
% That is, the read timestamp is chosen to be the maximum point
% at which the oplog of the primary has no gaps.
% This ensures the availability of the snapshot.

% In a sharded cluster, a mongos, as a transaction router,
% uses its local HLC as the read timestamp for the transaction.
% To address the snapshot unavailability problem,
% MongoDB will delay the transaction operations
% until the snapshot becomes available,
% as Clock-SI does~\cite{ClockSI:SRDS2013}.
%%%%%%%%%%%%%%%%%%%%
\subsubsection{YugabyteDB} \label{sss:yugabytedb}

YugabyteDB's distributed transaction architecture
is inspired by Google Spanner~\cite{Spanner:OSDI2012,Spanner:TOCS2013}.
% \footnote{YugabyteDB (DocDB transactions layer): https://docs.yugabyte.com/preview/architecture/transactions/}
Different from Spanner, YugabyteDB uses HLCs instead of atomic clocks.

Each write transaction is also assigned a commit hybrid timestamp.
Every read request in YugabyteDB is assigned a hybrid time,
namely the \emph{read hybrid timestamp}, denoted $\htread$.
$\htread$ is chosen to be the latest possible timestamp
such that all future write operations
{in the tablet} would have a strictly later hybrid time than that.
This ensures the availability of the snapshot.
%%%%%%%%%%%%%%%%%%%%
%%%%%%%%%%%%%%%%%%%%%%%%%%%%%%
% impl.tex

%%%%%%%%%%%%%%%%%%%%%%%%%%%%%%
\section{Lightweight Implementations of Timestamp-based Checking}
\label{section:impl}

\subsection{Additional Experimental Data}
\label{section:appendix-experiment}

%%%%%%%%%%%%%%%%%%%
% \subsection{Additional Experimental Data on \tool}
% \label{ss:appendix-experiment-offline}

% \input{figs/exp-decompose.tex}

% Figure~\ref{fig:decompose} shows the time consumption
% of each stage of \tool{} \emph{without} GC
% under varying workload parameters.

%%%%%%%%%%%%%%%%%%%
% \subsection{Additional Experimental Data on \onlinetool}
% \label{ss:appendix-experiment-online}
%%%%%%%%%%%%%%%%%%%%%%%%%%%%%

\begin{figure}[t]
	\centering
	\includegraphics[width = 0.45\textwidth]{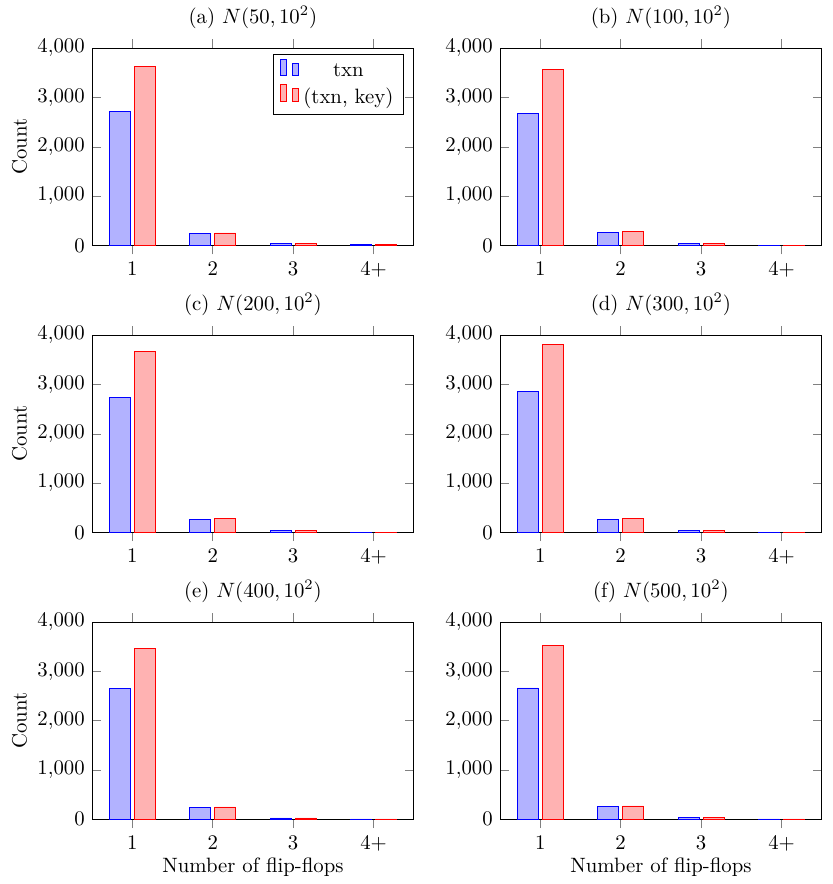}
	\caption{Statistics of flip-flops under varying delays.}
	\label{fig:online-flip-mu}
\end{figure}
\begin{figure}[t]
	\centering
	\includegraphics[width = 0.45\textwidth]{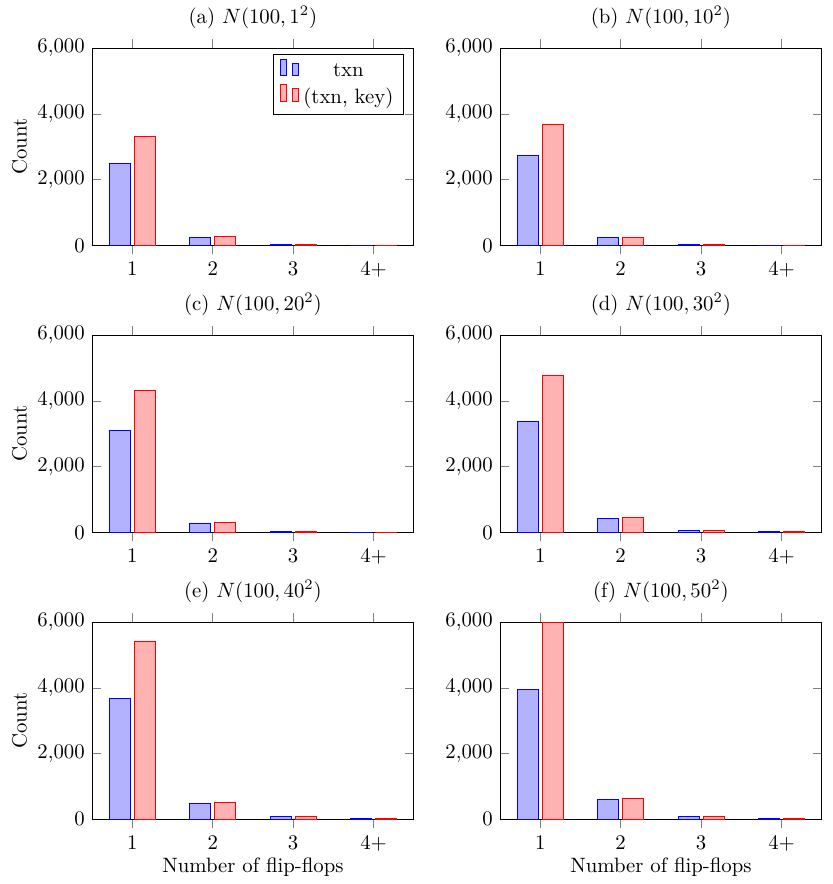}
	\caption{Statistics of flip-flops under varying standard deviations of delays.}
	\label{fig:online-flip-sigma}
\end{figure}
\begin{figure}[t]
	\centering
	\includegraphics[width = 0.45\textwidth]{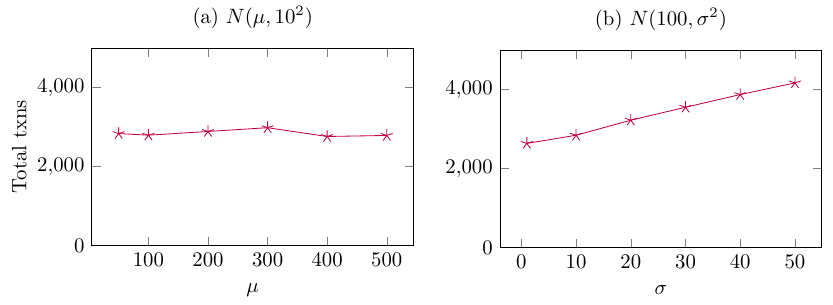}
	\caption{The number of unique transactions involved in flip-flops.}
	\label{fig:online-flip-total}
\end{figure}

\begin{figure}[t]
	\centering
	\includegraphics[width = 0.45\textwidth]{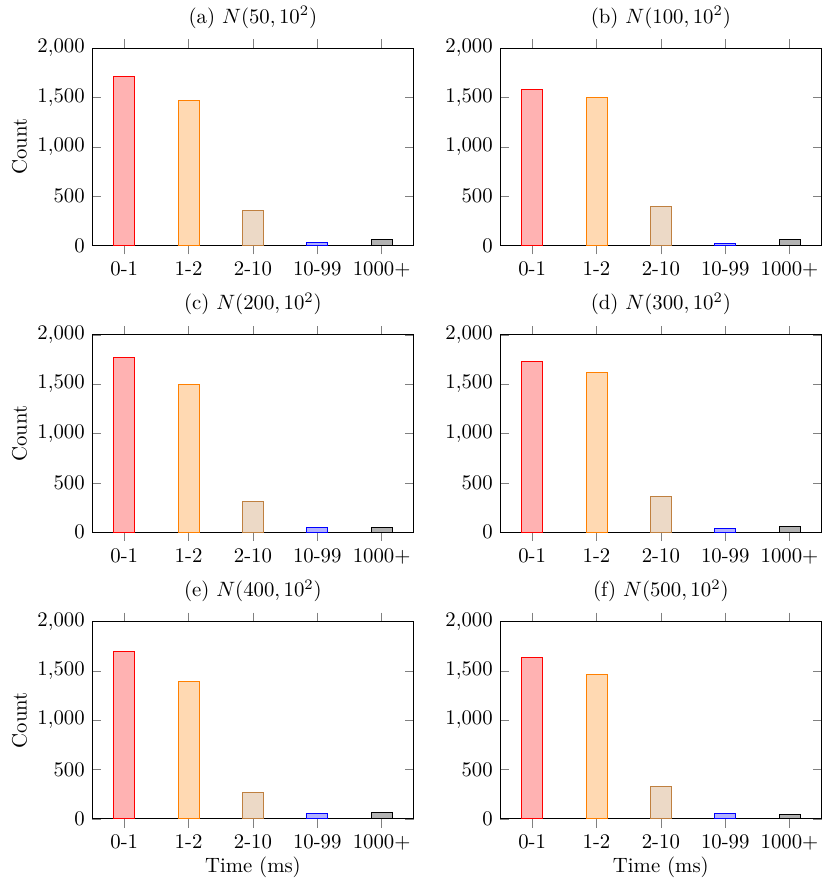}
	\caption{Time to finalize the result of checking \extaxiom{} under varying average delays.}
	\label{fig:online-flip-time-mu}
\end{figure}
\begin{figure}[t]
	\centering
	\includegraphics[width = 0.45\textwidth]{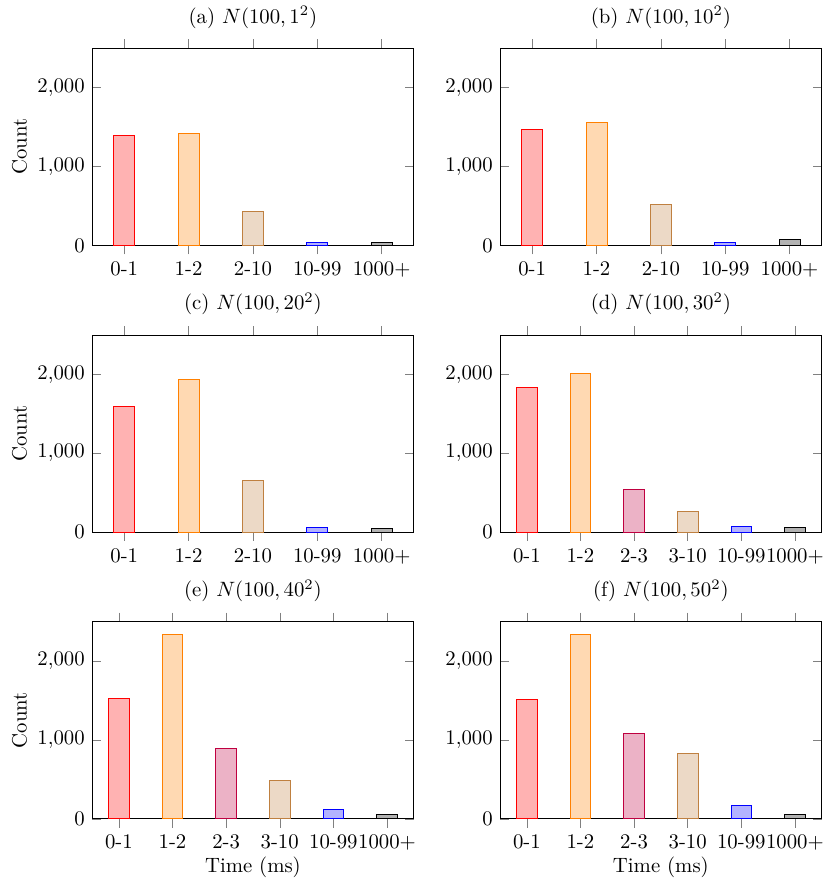}
	\caption{Time to finalize the result of checking \extaxiom{} under varying standard deviations of delays.}
	\label{fig:online-flip-time-sigma}
\end{figure}

In this part, we provide more experimental results with varying configuring settings.
% An \extaxiom{} violation manifests as a transaction-key pair.
In Figure~\ref{fig:online-flip-mu} and~\ref{fig:online-flip-sigma},
the right bar denotes the number of \extaxiom{} violations,
while the left bar denotes the number of
unique transactions (txn) involved in these violations.
In these experiments, we inject delays
following a normal distribution with a different mean value $\mu$
and a standard deviation of $\sigma$, i.e., $N(\mu,\sigma^2)$.
We observe that 20\% to 40\% of transactions exhibit flip-flops,
with the majority (99\%) experiencing them once or twice.
Furthermore, Figure~\ref{fig:online-flip-time-mu} and ~\ref{fig:online-flip-time-sigma} reveal that
over 95\% of the false positives/negatives are rectified within 10 ms.

Similar to the impact on flip-flops, there are similar conclusions regarding the number of transactions involved in these flip-flops, which are greatly influenced by the standard deviation, as shown in Figure~\ref{fig:online-flip-total}.

\begin{figure}[t]
	\centering
	\includegraphics[width = 0.42\textwidth]{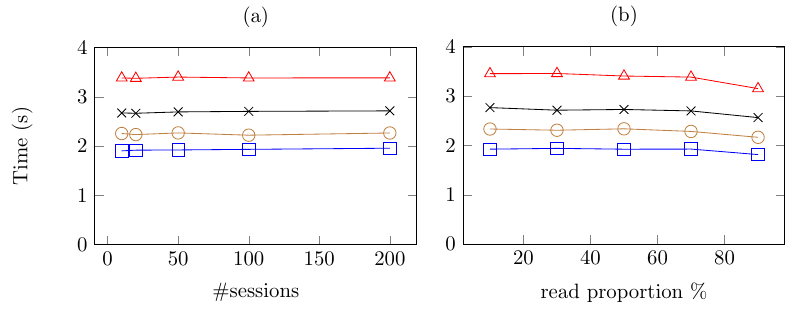}
	\caption{Runtime of \tool{} with various GC strategies
	  under varying workload parameters.}
	\label{fig:runtime11}
\end{figure}

\begin{figure}[t]
	\centering
	% \includegraphics[width = 0.30\textwidth]{figs/exp-online-tps-cobra.pdf}
	% \caption{Throughput of \onlinetool{} and Cobra over time.}
	% \label{fig:online-tps-cobra}
        \begin{subfigure}{0.235\textwidth}
		\centering
		\includegraphics[width = \textwidth] {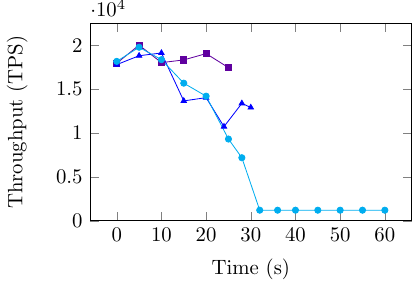}
		\caption{SI checking (RUBiS)}
		\label{fig:online-tps-rubis-si}
	\end{subfigure}
        \begin{subfigure}{0.235\textwidth}
		\centering
		\includegraphics[width = \textwidth] {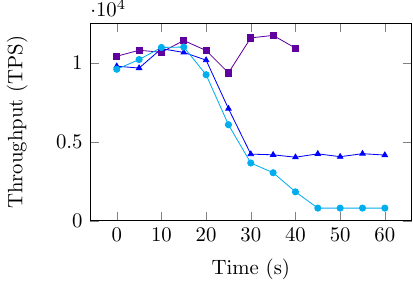}
		\caption{SI checking (Twitter)}
		\label{fig:online-tps-twitter-si}
	\end{subfigure}
    % \caption{\revise{Throughput of online checking over time.}}
    \caption{Throughput of online checking over time.}
    \label{fig:online-tps-throughput-rubis-twitter-si}
\end{figure}

% \revise{
To enhance the practicality of our research,
% in this revision,
we've incorporated the RUBiS and Twitter workloads. We integrated these workloads into our experiments to comprehensively evaluate our proposed online checker, \onlinetool{}. 
Figure~\ref{fig:online-tps-rubis-si} and Figure~\ref{fig:online-tps-twitter-si} show similar results of \onlinetool{} for different datasets to that of \onlinetool{}-SER. \onlinetool{} still outperforms Cobra in checking throughput against new datasets. 
% }

\begin{figure}[t]
	\centering
	% \includegraphics[width = 0.30\textwidth]{figs/exp-online-tps-cobra.pdf}
	% \caption{Throughput of \onlinetool{} and Cobra over time.}
	% \label{fig:online-tps-cobra}
        \begin{subfigure}{0.235\textwidth}
		\centering
		\includegraphics[width = \textwidth] {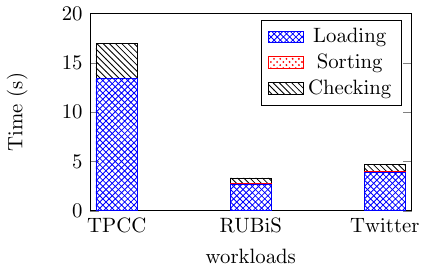}
		% \caption{SI checking (RUBiS)}
		% \label{fig:offline-tps-tpcc}
	\end{subfigure}
    % \caption{\revise{Throughput of offline checking for different workloads.}}
    \caption{Throughput of offline checking for different workloads.}
    \label{fig:offline-tps-tpcc}
\end{figure}

% \revise{
For TPC-C benchmark, We have evaluate \tool{} instead of \onlinetool{}.
The reason for this is that TPC-C involves numerous tables, most of which use composite primary keys, resulting in a very large range of primary key values. In online checking, maintaining the frontier corresponding to each timestamp requires significant computational resources, as evidenced by the comparison between Twitter and Rubis. In contrast, offline checking only needs to maintain a single global frontier, allowing it to easily handle any workload, including TPC-C. Figure \ref{fig:offline-tps-tpcc} shows the detailed evaluation of different workloads for \tool{}.
Note that Cobra is not constrained by this limitation; however, we contend that it does not qualify as a true online checking method. This is because it requires modifications to the user's workload (insert fence transaction \cite{Cobra:OSDI2020} split workloads into offline histories in batch) and halts upon detecting a violation.
% , as detailed in R4.O6.
% }

\begin{figure}[t]
	\centering
	% \includegraphics[width = 0.30\textwidth]{figs/exp-online-tps-cobra.pdf}
	% \caption{Throughput of \onlinetool{} and Cobra over time.}
	% \label{fig:online-tps-cobra}
        \begin{subfigure}{0.235\textwidth}
		\centering
		\includegraphics[width = \textwidth] {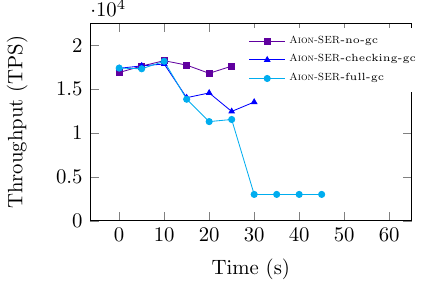}
		% \caption{SI checking (RUBiS)}
		% \label{fig:offline-tps-tpcc}
	\end{subfigure}
    % \caption{\revise{Throughput of online checking over time for non-conforming histories.}}
    \caption{Throughput of online checking over time for non-conforming histories.}
    \label{fig:online-tps-non-conforming}
\end{figure}

% \revise{
We also have conducted additional experiments using default benchmarks as shown in Figure \ref{fig:online-tps-non-conforming}.
We used the history generated under the SI level to test \onlinetool{}-SER. The experimental results show that \onlinetool{} detects all violations efficiently, with a checking speed comparable to that of histories without violations, while Cobra terminates immediately upon detecting the first violation. Additionally, we validated the number of violations using \tool{}-SER, confirming consistency with 11839 violations across 500K transaction runs.
% }

% \revise{
\subsection{Correctness of \onlinetool{}}
\label{section:appendix-proof}
The Aion algorithm follows a ``simulate-and-check'' approach, and its correctness is straightforward when all transactions are processed in ascending order based on their start/commit timestamps, ensuring no re-checking is required. However, when out-of-order transactions necessitate re-checking, it is crucial to determine which transactions should be re-checked against which axioms.
% }

% \revise{
Suppose that a transaction $T$ arrives out of order. We divide the set of previously processed transactions into three categories and, for each category (fixing a representative transaction $T'$), identify the set of axioms that should be re-checked. Clearly, the \intaxiom{} axiom for $T$' is not influenced by $T$ and thus does not need to be re-checked.
\begin{itemize}
    \item $T$' commits before $T$ starts: $T$' is not influenced by $T$, so no axioms need to be re-checked.
    \item $T$' commits after $T$ starts but starts before T commits: Since a transaction only considers transactions that committed before it as visible, the \extaxiom{} axiom for $T$' does not need to be re-checked. However, since $T$'overlaps with $T$, we must re-check the \noconflictaxiom{} axiom.
    \item $T$' starts after $T$ commits: The \extaxiom{} axiom needs to be re-checked, but the \noconflictaxiom{} axiom does not.
\end{itemize}
% }

% \revise{
% We have added the insight in Section \ref{ss:aion} and the formal argument to the appendix part of the technical report \cite{aion_and_chronos} due to space constraints. 
% }
%%%% appendix

% \input{}

\end{document}